\documentclass[journal]{IEEEtran}
\usepackage{amsmath,amsfonts}
\usepackage{algorithmic}
\usepackage{algorithm}
\usepackage{array}
\usepackage[caption=false,font=normalsize,labelfont=sf,textfont=sf]{subfig}
\usepackage{textcomp}
\usepackage{stfloats}
\usepackage{url}
\usepackage{verbatim}
\usepackage{graphicx}
\usepackage{cite}
\usepackage{bm}
\usepackage[switch]{lineno}
\usepackage{booktabs}
\usepackage{multirow}
\usepackage[colorlinks,
            linkcolor=red,
            anchorcolor=blue,
            citecolor=blue]{hyperref}
\usepackage{xcolor}
\usepackage{hyperref}

\newcommand*{\Beta}{\bm{\eta}}
\IEEEpubid{\begin{minipage}{\textwidth}\ \centering
		Copyright © 2025 IEEE. Personal use of this material is permitted. 
		However, permission to use this material for any other purposes must be obtained from the IEEE by sending an email to pubs-permissions@ieee.org.
\end{minipage}}
\begin{document}

\title{Reversible Unlearnable Examples: Towards the\\Copyright Protection in Deep Learning Era}

\author{Binze Wang, Jinyu Tian,~\IEEEmembership{Member,~IEEE}, Xingrun Wang, \\Xiaochen Yuan,~\IEEEmembership{Senior Member,~IEEE}, Jianqing Li,~\IEEEmembership{Senior Member,~IEEE}
\thanks{This research was partially supported by the National Natural Science Foundation of China (Grant No. 62202009), the Macau Science and Technology Development Fund (Grant Nos. 0040/2023/ITP1 and 0004/2023/RIB1), and the Basic and Applied Basic Research Foundation of Guangdong Province (Grant No. 2024A1515011755). (\textit{Corresponding author: Jinyu Tian.})
}
\thanks{B. Wang, J. Tian and J. Li are with the School of Computer Science and Engineering, Faculty of Innovation Engineering, Macau University of Science and Technology, Macau, China (e-mail: 3240007529@student.must.edu.mo; jytian@must.edu.mo; jqli@must.edu.mo).}
\thanks{X. Wang is with the School of Computer Science and Artificial Intelligence of FoShan University, FoShan 528225, China (e-mail:wangxingrun@fosu.edu.cn).}
\thanks{X. Yuan is with the Faculty of Applied Sciences of the Macao Polytechnic University, Macau, China (e-mail: xcyuan@mpu.edu.mo).}}

\markboth{Journal of \LaTeX\ Class Files,~Vol.~14, No.~8, August~2021}%
{Shell \MakeLowercase{\textit{et al.}}: A Sample Article Using IEEEtran.cls for IEEE Journals}


\maketitle

\begin{abstract}
Significant advancements in deep learning have been made possible by the utilization of large datasets, underscoring the critical importance of copyright protection. Adding meticulously designed perturbations to examples, making them unlearnable has become a crucial approach for safeguarding data copyright. Existing methods for creating unlearnable examples overlook the risk of data leakage, which can threaten data ownership. Thus, copyright protection in deep learning faces two main threats: illegal model training and malicious data leakage. We investigate that these two threats cannot be solved by straightforwardly combining existing availability attacks and watermarking techniques as their negative interaction effects. Therefore, in this paper, we propose a novel copyright protection mechanism for the aforementioned security concerns. Considering that the prevention of unauthorized model training requires powerful generalizability of unlearnable perturbations, we generate perturbations to induce the model to learn uncorrelated features of input images. It works by minimizing the mutual information of the input and output of the model. On the other hand, to eliminate the side impact of unlearnable perturbations on the watermark extraction, we design a dual extraction strategy by using two distinct watermark extractors. Extensive experiments on the image datasets {ImageNet, CIFAR10, and Pets} show that our proposed method could provide comprehensive copyright protection to images. The code is available at \href{https://github.com/Yeah21/ReversibleUnlearnableExamples}{https://github.com/Yeah21/ReversibleUnlearnableExamples}.
\end{abstract}

\begin{IEEEkeywords}
Copyright protection, reversible unlearnable examples, watermarking technique.
\end{IEEEkeywords}

\section{Introduction\label{sec:intro}}
\IEEEPARstart{C}{opyright} protection has always been one of the key issues in the development of human civilization. As we enter the Internet era, digital image watermarking plays a vital role in copyright protection by validating image ownership against widespread {unauthorized manipulation} \cite{10.1145/3711930} and leakage on online platforms \cite{ernawan2023recent}.
This technique operates by embedding identification information into images without compromising their usability and enables \textbf{ownership verification} by extracting watermarks from the protected images \cite{10520319,9343885,9374479,8708267,9928284}.
\begin{figure}
  \vspace{-1.0em}
    \centering
\includegraphics[width=0.42\textwidth]{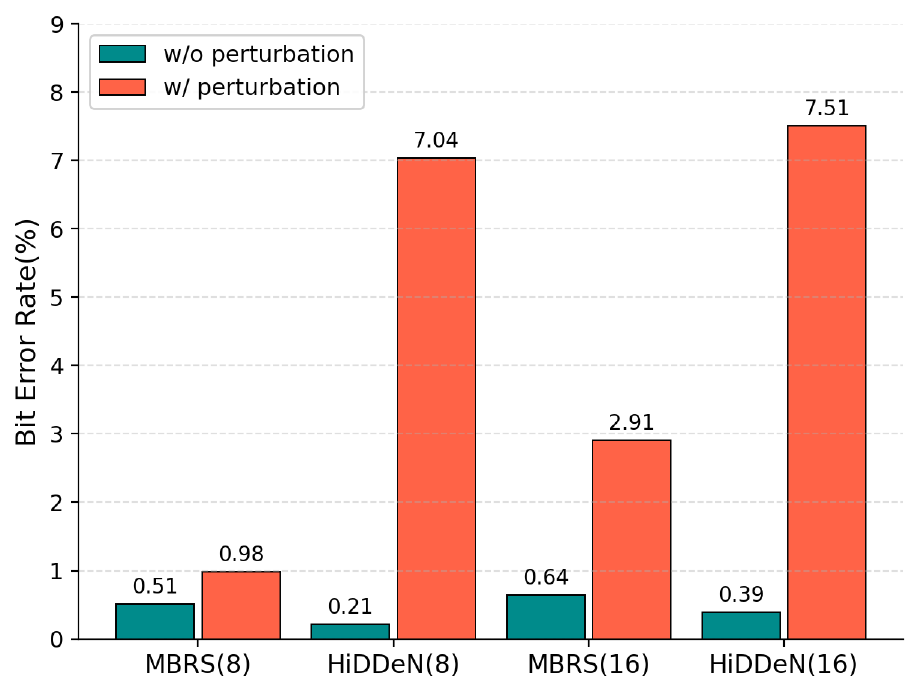}
\vspace{-1.0em}
  \caption{{Impact of unlearnable perturbations on watermark extraction performance (Bit Error Rate$\downarrow$).}  MBRS \cite{jia2021mbrs} and HiDDeN \cite{zhu2018hidden} are two watermark embedding methods. The value in the bracket is the budget of unlearnable perturbation.}
  \label{fig1}
\vspace{-1.0em}
\end{figure}

{In the era of deep learning, digital watermarking techniques have expanded their applicability to emerging challenges, including safeguarding model copyrights \cite{9833747} and preventing unauthorized model replication \cite{10.1145/3664647.3685507}.} 
However, a formidable new challenge has emerged with the swift proliferation of deep learning techniques. The risk comes up when valuable images might be taken by unauthorized users to train commercial models, seriously hurting the rights of the image owners \cite{hill2022secretive,zhang2020adversarial}. These leaked images, typically hidden from public sight, present an insurmountable barrier to traditional watermarking techniques for copyright verification. As a result, protecting the copyright of images should go beyond just confirming ownership and include preserving their intrinsic semantic information to train deep learning models. For convenience, we regard the second type of copyright protection as \textbf{semantic information protection}. A commonly employed approach to protect semantic information involves the introduction of imperceptible adversarial perturbations into the protected images. This strategy effectively thwarts models to learn valuable semantic features from the images, which is often referred to as the \textit{unlearnable examples }\cite{fowl2021adversarial,sandoval2022autoregressive,yu2022availability,peng2022learnability,fowl2021preventing,fu2022robust,huang2021unlearnable,zhang2023unlearnable}. 

\IEEEpubidadjcol
The two categories of copyright protection issues include ownership verification and semantic information protection, and there are perfect solutions for each type of problem. However, developing a comprehensive copyright protection mechanism that effectively addresses both challenges is undeniably intricate. The reason is that both digital watermarks and unlearnable perturbations subtly alter images, and when the two techniques are combined, the unlearnable perturbations would complicate the extraction of watermark information. For example, as shown in Fig. \ref{fig1}, when the unlearnable perturbations are involved, the performance of the watermarking extraction is degraded significantly, manifested by the increase in Bit Error Rate (BER) from $0.39\%$ to $7.51\%$ (the rightmost bar). This observation shows a straightforward combination of watermarking and unlearnable example methods cannot effectively deal with both ownership verification and semantic information protection at the same time. Therefore, achieving a harmonious balance between the two objectives requires innovative approaches that can mitigate their interference and achieve a mutually beneficial coexistence.

Indeed, current approaches have explored combining watermark information with unlearnable perturbation, although they may not directly address the challenge of interference between the two. The GEAA method \cite{zhao2022guided} tackles the problem by embedding watermark information while removing unlearnable perturbation to avoid interference-related problems. {This design restricts watermark extraction to recovered data only, failing to verify ownership for perturbed images.} The Adv-watermark \cite{jia2020adv} utilizes a visible watermark as a small fixed noise element and achieves an evasion attack effect by adjusting the watermark's position and transparency within the image. However, visible watermarks are very easy to attack \cite{10464320}, resulting in protection failure, and the evasion attack performs unsatisfactorily in the task of availability attack. 

\begin{figure*}
  \centering
  \includegraphics[width=0.95\textwidth]{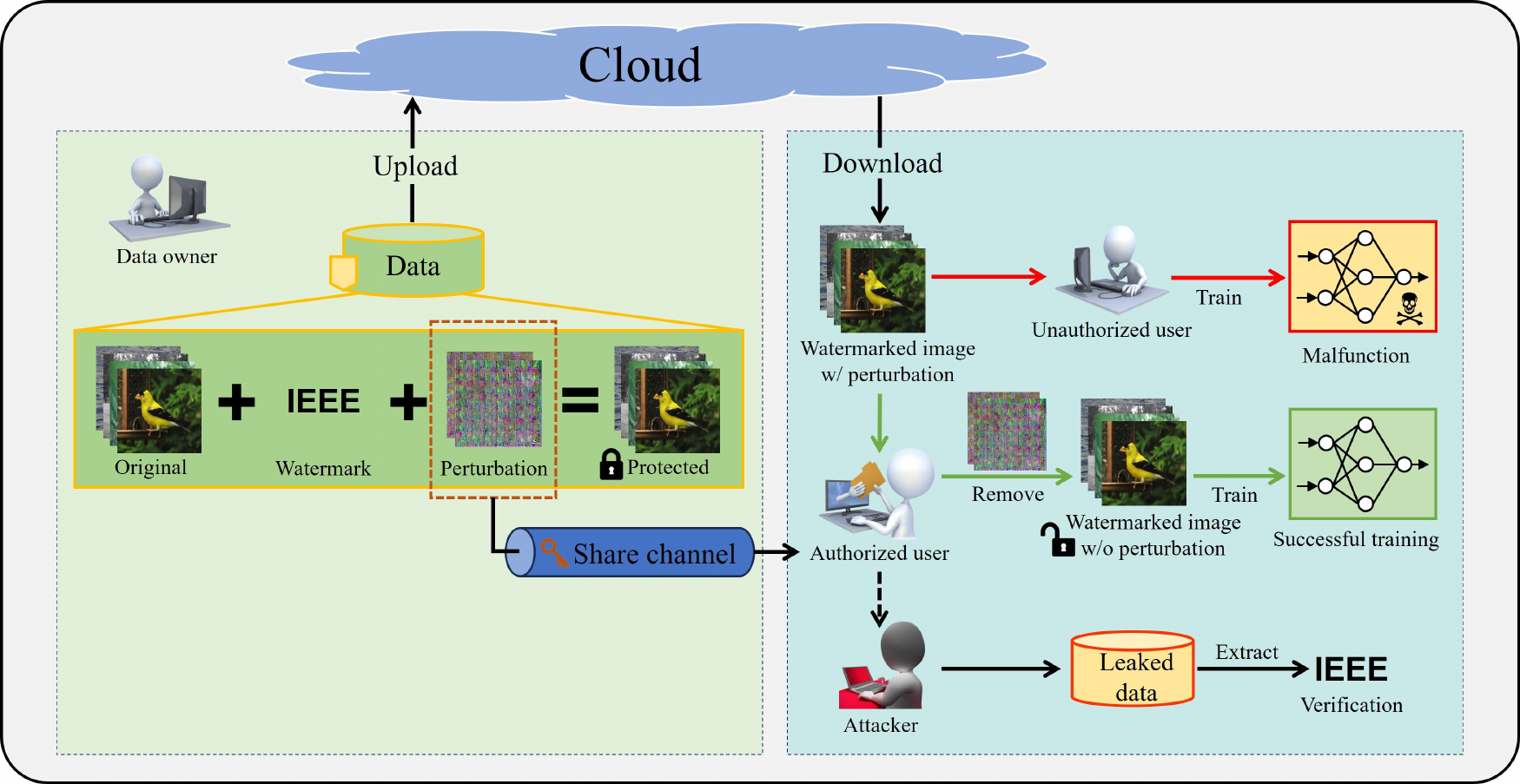}
  \caption{{Application scenario involving data owner, authorized user, unauthorized user, and attacker.}}
  \label{fig-Applicationscenario}
\vspace{-1.0em}
\end{figure*}
In this paper, we present a novel protective mechanism designed to address the emerging challenges in copyright infringement within the context of the deep learning era. Our proposed mechanism facilitates the concurrent utilization of watermarking and unlearnable perturbation. Each part functions normally without compromising the effectiveness of the other. {Specifically, we focused on a real-world scenario illustrated in Fig. \ref{fig-Applicationscenario}, involving four key roles: the data owner, authorized users, unauthorized users, and a potential attacker. The data owner curates a high-quality dataset and intends to make it publicly accessible while retaining verifiable ownership and preventing unauthorized exploitation. To this end, the data owner takes preventive actions by injecting both watermarks and imperceptible perturbations (the left square region). These perturbations render the semantic information within the data unlearnable by machine learning models, acting as a cryptographic mechanism.} Subsequently, these protected images are uploaded to cloud services like Google Drive, enabling public access and download. The embedded unlearnable perturbations play a vital role in effectively deterring unauthorized users' misuse for training commercial deep models, serving the goal of semantic information protection (the solid red line). 
On the flip side, when authorized users access the protected images, they can legally train their deep models by removing the perturbations acquired through a secure channel (the solid green line). This requires the designed unlearnable perturbations to be reversible. 
{The class-wise perturbation design inherently ensures that perturbations are label-dependent and uniformly applied across all images within a class, enabling efficient recovery.
However, an attacker may steal decryption keys (e.g., via compromised channels or illicit purchase) to remove perturbations, then redistribute clean data, violating both copyright and accessibility control (the dashed black line). Therefore, the watermark should work normally, regardless of whether there are perturbations in the images.} Our work tackles the issue by introducing a dual watermarking extractor, which significantly mitigates the interference caused by the unlearnable perturbation, and ensures that we can extract the same watermark successfully with or without unlearnable perturbations.  

In summary, our work concentrates on the crucial modules in the scenario described above, and derives the following novelties and contributions:
\begin{itemize}
  \item We propose a practical and comprehensive copyright protection framework for image datasets. This innovative framework employs embedded unlearnable perturbations within protected images, thwarting unauthorized users from training a deep model. Significantly, this framework also pioneers a comprehensive ownership verification approach, ensuring that watermarks can be successfully extracted whether the interference of unlearnable perturbations is present or absent.

  \item We propose a novel methodology to generate unlearnable perturbations from the perspective of mutual information. Our motivation revolves around the viewpoint that an effective unlearnable perturbation could push the output of a model independently with the input of this model. Naturally, the model cannot learn useful information from the unlearnable images.    
  
  \item To enhance the convenience and feasibility of our copyright protection framework, firstly, we develop a class-wise universal unlearnable perturbation (it means that images of the same class share the same unlearnable perturbation), allowing for a low-overhead recovery process of semantic information in images. Moreover, we incorporate the widely employed image preprocessing technique, JPEG compression, into our framework, which significantly enhances the robustness of the watermarks and unlearnable perturbations.   
  
  \item Empirical evaluations conducted on public datasets show that our approach achieves the lowest average accuracy on unauthorized model training and simultaneously manifests a commendable watermark extraction performance. These experiment results demonstrate that the proposed framework achieves the two predefined goals of copyright protection: namely, ownership verification and semantic information protection. 
\end{itemize}

The rest of this paper is organized as follows: Section \ref{RelatedWorks} reviews the related works on unlearnable perturbation and digital watermarking. Section \ref{ProposedMethod} presents robustness of perturbations and watermark verification. Experimental results are given in Section \ref{Experiments} and Section \ref{Conclusion} concludes the paper. 
The main notations used throughout this paper with their definitions can be found in Table \ref{tab:notations}.

\begin{table}
    \vspace{-1.0em}
    \centering
    \caption{Summary of Notations}
    \begin{tabular}{ m{4em}  m{6cm}}
    \hline
    \rule{0pt}{8pt}Notation     &\rule{0pt}{8pt}Description  \\ \hline
    \rule{0pt}{8pt}$\bm{x}_i$ &\rule{0pt}{8pt}original image \\
    $\tilde{\bm{x}}_i$ &watermarked image \\
    $\hat{\bm{x}}_i$ &protected image (image with watermark and unlearnable perturbation)\\
    $\bm{m}$ &watermark \\
    $E$ &encoder \\
    $D$ &decoder \\
    $\bm{\eta}_c$ &class-wise unlearnable perturbation \\
    $C$ &number of classes \\
    $\epsilon$ &budget (the limitation of perturbation) \\
    $f$ &the victim model for perturbation generation \\
    $\mathcal{J}$ &JPEG compression \\ 
    $\mathcal{D}$ &original dataset \\
    $\tilde{\mathcal{D}}$ &watermarked dataset \\
\hline
    \end{tabular}
    \label{tab:notations}
\vspace{-1em}
\end{table}

\section{Related Works\label{RelatedWorks}}
\subsection{Unlearnable Perturbation}
Generally, the data availability attack involves adding perturbations to training data, leading to networks malfunctioning on the test data. 
It is a bi-level optimization problem \cite{DBLP:conf/icml/BiggioNL12} by minimizing the victim model's cross-entropy loss on the training data while maximizing its classification error on clean test data. Although direct optimization of the objective is intractable for deep neural networks \cite{munoz2017towards}, there are some approximate solutions based on gradient matching\cite{DBLP:conf/nips/HuangGFTG20,geiping2020witches,fowl2021preventing}, and Neural Tangent Kernels\cite{DBLP:conf/icml/YuanW21}.
To simplify the problem, simple optimization objectives are proposed, including error-minimizing noise \cite{huang2021unlearnable,peng2022learnability,fu2022robust} and error-maximizing noise \cite{fowl2021adversarial}. The former perturbations are optimized to reduce the training loss to zero then the target model will have nothing to learn from the data. The latter shows that the common adversarial examples are sufficient to generate powerful unlearnable perturbations. {TUE \cite{rentransferable} investigated transferable unlearnable examples, improving the robustness and cross-model applicability of perturbations through optimized design. Besides the classification task, UnSeg \cite{sun2024unseg} is a universal unlearnable example generator designed for image segmentation tasks.}

In order to generate perturbations more quickly, \cite{sandoval2022autoregressive} proposed a method of generating perturbations by a linear filter, which does not require any optimization process. Similarly, \cite{yu2022availability} unveils a linear-separability property of unlearnable noise and proposes the Synthetic Perturbations method to directly synthesize perturbations as effective unlearnable noise.  

\subsection{Digital Watermark}
Digital image watermarking aims to embed secret messages into images and then extract them for verification. In recent years, many DNN-based methods \cite{ahmadi2020redmark,liu2019novel} have been applied in watermarking for strong robustness.
For example, HiDDeN \cite{zhu2018hidden} employs an encoder-decoder-based framework for watermarking.
StegaStamp \cite{tancik2020stegastamp}  focuses on print-shooting watermarks extraction, incorporating several differential attack methods into the noise layer. A two-stage separable framework \cite{liu2019novel} is proposed to deal with nondifferentiable operations in the noise layer.
JPEG compression is a commonly used method for reducing the file size of images, particularly for storage and transmission purposes. MBRS \cite{jia2021mbrs} specifically aims to enhance the robustness of the watermark against JPEG compression artifacts and introduces a random noise layer containing three different JPEG noise types into their network. 
Sun et al. \cite{9103635} proposed a novel robust image watermarking method in the DCT domain, specifically tailored for embedding and extracting watermarks in images shared on online social networks.
{Recent advances further explore robustness in screen-shooting scenarios. WaveRecovery \cite{10772578} leverages wavelet domain encoding and noise recovery decoding to enhance visual quality and extraction accuracy, while \cite{10559271} simulates perspective-dependent grayscale deviations (GDF) to improve imperceptibility.}
\begin{figure*}
  \centering
  \includegraphics[width=0.9\textwidth]{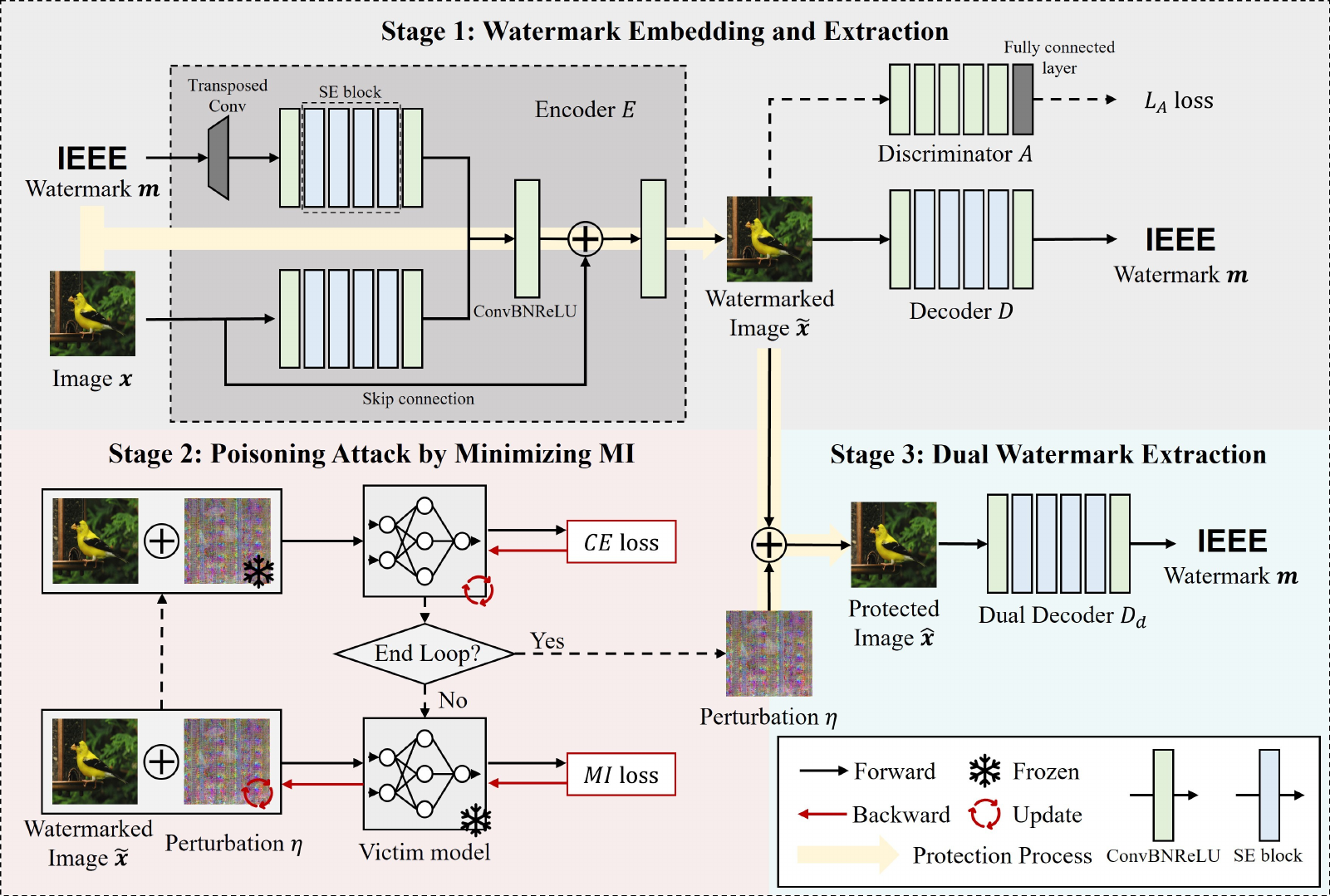}
  \caption{{The pipeline of the proposed copyright protection framework.}}
  \label{figure11}
\vspace{-1.0em}
\end{figure*}
\section{Proposed Method\label{ProposedMethod}}
In this paper, we redefine the goals of copyright protection for data in the deep learning era, including \textbf{ownership verification} and \textbf{semantic information protection}. Recalling the description of Fig. \ref{fig-Applicationscenario}, in our considered application scenario, the data owner's images face two different aspects of copyright infringement. On one hand, unauthorized users seek to employ these images for training commercial deep models, leading to the infringement of semantic information. {On the other hand, the attacker maliciously leaks unprotected data which are without unlearnable perturbations.} In this scenario, we need to embed unlearnable perturbations to prevent unauthorized training and watermarks to verify the ownership of images. The watermarks in the protected images should be extractable both before and after the removal of the perturbations. This unique feature sets our proposed copyright protection framework apart from existing studies. 
We now delve into the comprehensive pipeline of our methodology.

\subsection{Outline of the Proposed Method}

The scheme of our copyright protection methodology follows the diagram in {Fig. \ref{figure11}}, {which consists of the watermark embedding and extraction} (\textcolor[HTML]{595959}{upper half}), {the unlearnable perturbation generation process} (\textcolor[HTML]{EC93AD}{lower left half}), {and dual watermark extraction} (\textcolor[HTML]{ABDDDC}{lower right half}). {In the watermark embedding and extraction part, the encoder $\bm{E}$ and decoder $\bm{D}$ are trained to process the watermark. Specifically, the encoder embeds a given watermark $\bm{m}$ into the original images $\bm{x}_i$'s and results in their watermarked counterparts $\tilde{\bm{x}}_i$'s. Correspondingly, the decoder extracts the watermark information $\bm{m}$ from the watermark images $\tilde{\bm{x}}_i$'s. Furthermore, to enhance the watermarked image quality, a discriminator is incorporated during training, enabling adversarial optimization. Unlearnable perturbation generation is an iterative optimization process as follows. Firstly, fix the perturbations and minimize the cross-entropy loss to update the victim model parameters. Secondly, fix the victim model parameters and minimize mutual information objectives to optimize the perturbations. These two steps are repeated until convergence, resulting in class-wise unlearnable perturbations.} As we emphasized before, our proposed framework has the unique feature of verifying the ownership of images both with and without the interference of the unlearnable perturbation. We should extract watermark $\bm{m}$ from both the watermarked image $\tilde{\bm{x}}$ and the protected image $\hat{\bm{x}}$. Note that the decoder $\bm{D}$ is jointly trained with the encoder $\bm{E}$ over original images $\bm{x}$'s. It generally can only extract the watermark from the watermarked image $\tilde{\bm{x}}_i$. Therefore, we design a dual decoder $\bm{D}_d$ to extract a watermark from the protected image $\hat{\bm{x}}_i$, which has an unlearnable perturbation $\bm{\eta}_c$ as the interference factor compared with $\tilde{\bm{x}}_i$. We call the designed encoder and decoder structure dual watermark extraction, as two decoders seek to reveal the same information from a common encoder.  
{In the {Fig. \ref{figure11}}, we highlight the process of obtaining protected images with perturbations and watermarks from clean images with yellow thick arrows. Specifically, for each image $\bm{x}_i$ of the $c$-th class, we obtained its protected version $\hat{\bm{x}}_i$ by first embedding watermark $\bm{m}$ using the encoder $\bm{E}$ and then adding the $c$-th class perturbation $\bm{\eta}_c$ on it, i.e., $\hat{ \bm{x}}_i = \bm{E}(\bm{x}_i) + \bm{\eta}_c$. When ownership verification is required, given the undetectability of imperceptible perturbations $\bm{\eta}_c$, we deploy dual watermark extractors operating in parallel. Decoder $\bm{D}$ processes the input image under the perturbation-free assumption, yielding extracted watermark $\bm{m}'$. Dual Decoder $\bm{D}_d$ processes the same input under the perturbation-present assumption, outputting $\bm{m}''$. Ownership is established if either extracted watermark matches the original embedded watermark $\bm{m}$.} 

For generating the class-wise unlearnable perturbations $\Beta_c$'s, we alternately train a victim model $f$ and search the perturbations $\Beta_c$'s. The training process is a competition between the model $f$ that aims to learn semantic information from the protected images $\hat{\bm{x}}_i$'s and the unlearnable perturbations $\Beta_c$'s that struggle to impede the learning process of model $f$. The model generally fulfills the learning process by minimizing the classification error, typically the cross-entropy. To impede the learning process of model $f$, we propose a novel \textbf{Unlearnable Examples by Minimizing the Mutual Information (MI)} between the model input and output. Since unauthorized models usually differ from the ones used to create unlearnable examples, these examples need to have good transferability. We will explore the proposed MI in Section \ref{poisoning}.  

We alternate training watermarks and perturbations. First, a network embeds and extracts watermarks, generating watermarked images. Then, we apply unlearnable perturbations to the watermark image. Finally, the dual watermark extractor is trained on images with both perturbation and watermark. Now, we discuss the watermark generation process and the unlearnable perturbation generation process, respectively.

The training process for generating watermarks and unlearnable perturbations involves several steps. First, we train a network with an encoder and decoder to produce watermarked images. Next, we apply unlearnable perturbations to these watermarked images using our proposed MI method. Finally, we train the dual watermark extractor on images with both perturbations and watermarks. In the following sections, we will delve into the details of how watermarks and unlearnable perturbations are generated.

\subsection{Watermark Embedding and Extraction\label{WatermarkEmbedding andExtraction}}
At present, the digital watermark method based on the neural network commonly contains an encoder and a decoder. The encoder $\bm{E}$ inputs watermark information $\bm{m}$ and the original image $\bm{x}_i$, outputs watermarked image $\tilde{\bm{x}}_i$; The decoder $\bm{D}$ extracts the watermark information $\bm{m}'$ from $\tilde{\bm{x}}_i$. The objective of optimizing network parameters is to ensure high image quality and high accuracy of watermark extraction. To evaluate the image quality, a straightforward selection is the Mean Square Error (MSE) between the watermarked image $\tilde{\bm{x}}_i$ and the original image $\bm{x}_i$. However, the work \cite{zhang2018unreasonable} points out that the similarity perceived by human eyes is not completely consistent with the pixel-level similarity index, and it is more appropriate to use semantic feature extraction by the networks. Therefore, we use a discriminant network to ensure the visual effect further. As for the extraction of watermark information, we treat this task as a binary classification problem where the watermark extraction network should correctly predict each bit of the watermark. This suggests we measure the difference between the embedded and the extracted watermarks using Binary Cross Entropy (BCE). To sum up, the loss of watermark embedding is as follows: 
\begin{linenomath*}
\begin{equation}
\begin{split}
\label{loss1}
\mathcal{L}_1&=\text{MSE}(\tilde{\bm{x}}_i-\bm{x}_i)+\log(\bm{A}(\tilde{\bm{x}}_i))+\text{BCE}(\bm{m},\bm{m}'),
\end{split}
\end{equation}
\end{linenomath*}
where
\begin{linenomath*}
\begin{equation}
\begin{split}
\tilde{\bm{x}}_i=\bm{E}(\bm{x}_i,\bm{m};\bm{\theta}_{\bm{E}}),~\bm{m}'=\bm{D}(J(\tilde{\bm{x}}_i);\bm{\theta_D}),
\end{split}
\label{loss2}
\end{equation}
\end{linenomath*}
$\bm{\theta_E}$, $\bm{\theta}_{\bm{D}}$ are parameters of the encoder and decoder, respectively, and $\bm{A}$ represents the discriminator. Here, we involve a differentiable JPEG module $J$ \cite{10.1145/103085.103089} to enhance the robustness of watermark extraction against the compression preprocess.  

Our watermark embedding and extraction modules are based on the MBRS \cite{jia2021mbrs}.
{The message is reshaped and amplified through a $3\times3$ convolution layer, features are expanded and extracted using transposed convolution layers and Squeeze-and-Excitation (SE) block \cite{Hu2017SqueezeandExcitationN}.}
In the encoding process, the cover image undergoes similar amplification and feature extraction, and the resulting features from both the message and cover image are concentrated and mapped through convolutional layers to produce the watermarked image. The decoding process entails amplifying a watermarked image, channel-wise feature conversion, and reshaping to retrieve the decoded message. 
{The discriminator consists of five convolutional blocks followed by a fully connected layer. Each block includes a 3$\times$3 convolution layer, Batch normalization, and ReLU activation. Fully connected layer with output dimension 1, using sigmoid activation to predict whether the input is a watermarked or original image.} 
Now, we can obtain the watermark encoder $\bm{E}$ and decoder $\bm{D}$ by minimizing the integration of $\mathcal{L}_1$. 
A dual watermark extractor is mentioned in Section \ref{dualwatermarkextraction} to extract watermarks from images with perturbation better.

\subsection{The Generation of Unlearnable Perturbations} \label{poisoning}
Upon having watermarked images $\tilde{\bm{x}}_i$'s, we then generate class-wise unlearnable perturbations $\bm{\eta}_c$'s to obtain the protected images $\hat{\bm{x}}_i$'s.
In our proposed copyright protection framework, we propose a novel information theory-based unlearnable example method. 
{We observe that existing methods \cite{fowl2021preventing,fowl2021adversarial,peng2022learnability,sandoval2022autoregressive,yu2022availability} highly rely on the classification information of the victim model, thereby causing the classification results to become inconsistent with the true labels.
Geometrically, perturbations shift normal samples in the direction normal to the decision boundary, enabling minimal alterations to cross the boundary \cite{goodfellow2014explaining,moosavi2016deepfool}. However, since the decision boundary varies significantly across different models \cite{8103145,liu2017delving}, this results in poor generalization of perturbations.
Consequently, error-maximization strategies generate perturbations tailored exclusively to the decision boundaries of a specific well-trained victim model.}

\begin{figure*}
  \centering
  \includegraphics[width=0.8\textwidth]{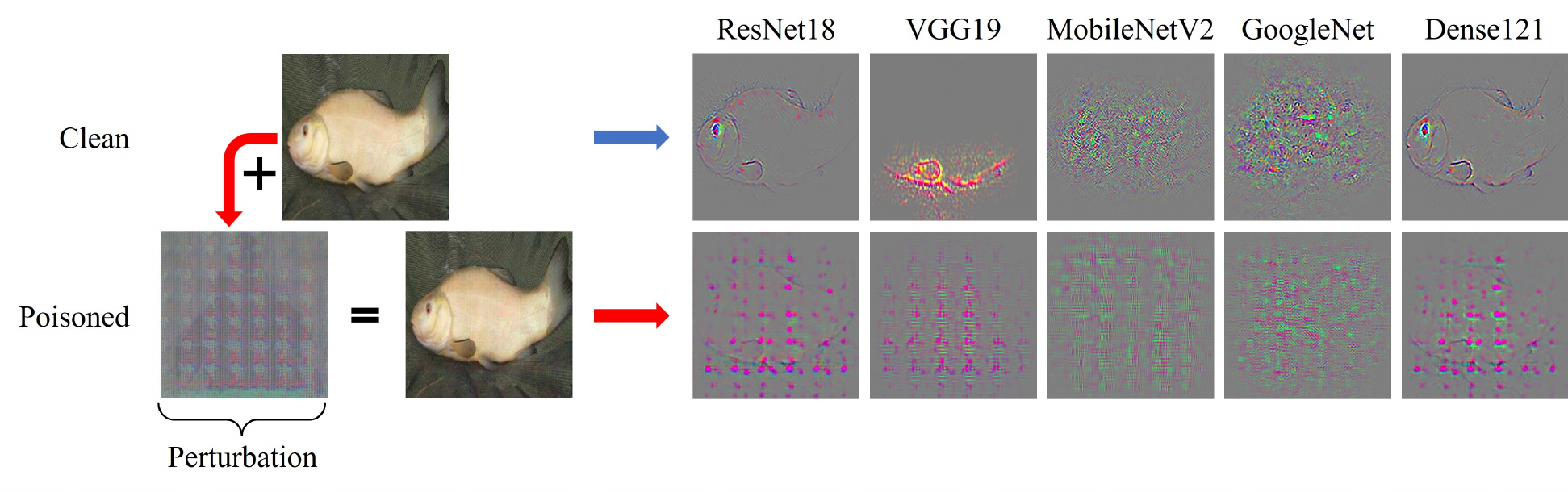}
  \vspace{-0.5em}
  \caption{{The CAM visualization of the clean model and the poisoned model.}}
  \label{fig:CAM}
\vspace{-0.5em}
\end{figure*}

{We overcome above limitation by exploiting the commonalities observed across models during training. Specifically, existing research indicates that different models tend to extract similar features from the same image dataset \cite{bansal2021revisiting,WOS:000922928202068,kornblith2019similarity,morcos2018insights,nguyen2020wide}. 
As shown in the 1-st row of {Fig. \ref{fig:CAM}}, our CAM visualizations reveal that diverse models rely on identical semantic features (such as fish texture and contour information).
This inspires us that if a deliberately crafted perturbation added to an image can dominate the features extracted by models, it will disrupt their ability to capture semantic information, thus resulting in a transferable poisoning effect.}

{To achieve effective intervention in the dependency between image semantic information and labels, a suitable metric is required to serve as the optimization objective. Mutual information (MI) is an ideal measure for quantifying the statistical dependence between two random variables. Intuitively, it represents the reduction in uncertainty of one variable when the other is known. The MI is calculated as} 
\begin{equation}
{I(X;Y)=H(Y)-H(Y|X),}
\end{equation}
{where the entropy $H(Y)$ measures the inherent uncertainty of variable $Y$, and the conditional entropy $H(Y|X)$ captures the remaining uncertainty in $Y$ given knowledge of $X$. A smaller MI $I(X;Y)$ indicates that even when $X$ is known, $Y$ remains highly uncertain. This aligns perfectly with our objective: we aim to introduce a perturbation $\eta$ to obscure the semantic information in images $X$, preventing the model from establishing a reliable association between $X$ and the label $Y$. 
From an information-theoretic perspective, perturbing the image to $X+\eta$ increases the difficulty for the model to determine the label, which is reflected in an increase in conditional entropy: 
$I(X+\eta;Y)=H(Y)-H(Y|X+\eta)<H(Y)-H(Y|X)=I(X;Y)$. Thus, we adopt a mutual information minimization strategy to weaken the dependence between image features and labels, pushing the learned features to be uncorrelated with the image semantic information.
The 2-nd row of {Fig. \ref{fig:CAM}} demonstrates that when perturbations (ResNet18 as the victim model) are added, all models shift attention exclusively to the perturbed regions, fundamentally disrupting semantic feature extraction.
This prevents any model from extracting features correlated with semantic content, making the data fundamentally "unlearnable". }

Formally, suppose we have obtained the watermarked versions $\tilde{\bm{x}}_i$'s of all training images $\bm{x}_i$'s. Let $\bm{\eta}_i$'s be the unlearnable perturbation corresponding to each $\tilde{\bm{x}}_i$. Considering the output $\hat{\bm{y}}_i$ of the victim model $f$ with respect to the input $\hat{\bm{x}}_i$, i.e., $\hat{\bm{y}}_i = f(\hat{\bm{x}}_i)$, we have the set $\hat{Y}$ of the outputs $\hat{\bm{y}}_i$'s. According to the discussion above, we can find unlearnable perturbations $\bm{\eta}_i$'s by minimizing the mutual information between $\hat{X}$ and $\hat{Y}$. That is, 
\begin{linenomath*}
\begin{equation}
\begin{aligned}
\label{IB_principle}
&\min \limits_{\bm{\eta}_i} I(\hat{X};\hat{Y}), \quad s.t. \quad \| \bm{\eta}_i \|_{\infty}<\epsilon, \\ \end{aligned}
\end{equation}
\end{linenomath*}
where $\hat{X}$ is parameterized by the optimization variables $\bm{\eta}_i$'s.
{Following with equation (9) in the reference \cite{zhao2020maximum}, we can derive the subsequent procedure. According to the definition of mutual information, $I(\hat{X};\hat{Y})=H(\hat{Y})-H(\hat{Y}|\hat{X})$, where $H(\hat{Y})$ is the entropy of model output $\hat{Y}$, and $H(\hat{Y}|\hat{X})$ is the conditional entropy of output given input $\hat{X}$. In deterministic models (like neural networks), the output $\hat{Y}$ becomes deterministic when input $\hat{X}$ is given, i.e., $H(\hat{Y}|\hat{X})=0$. Therefore, followed with \cite{zhao2020maximum}, we can derive $I(\hat{X};\hat{Y})=H(\hat{Y})$. The optimization objective simplifies to}

\begin{linenomath*}
\begin{equation}
\label{IB_principle1_quote}
{\min \limits_{\bm{\eta}_i} H(\hat{Y}), \quad s.t. \quad \| \bm{\eta}_i \|_{\infty}<\epsilon.}
\end{equation}
\end{linenomath*}
{$H(\hat{Y})=-\sum_j(p_j\log{p_j})$, where $p_j$ is the model's predicted probability for the $j$-th class. Based on the above derivation, we can reformulate Equation (\ref{IB_principle}) as:}
\begin{linenomath*}
\begin{equation}
    \min \limits_{\|\ \bm{
    \eta}_i\|_{\infty}<\epsilon} \mathbb{E}_{ {\tilde{\bm{x}}}_{i} \in \tilde{\mathcal{D}}} \{ -f(J(\tilde{\bm{x}}_i+{\bm{\eta}}_i)) \log f(J(\tilde{\bm{x}}_i+{\bm{\eta}}_i)) \}.
\label{PAIB2}
\end{equation} 
\end{linenomath*}
Here we also involve a differentiable JPEG module \cite{10.1145/103085.103089} as the generation of watermarks to enhance the performance of the unlearnable perturbations in practice. To solve the problem (\ref{PAIB2}), we adopt the first-order method as follows:
\begin{linenomath*}
\begin{equation} \label{JEPG_I}
 \Beta_i = -l_\eta\nabla_{\tilde{\bm{x}}_i} \mathcal{L}_p(\tilde{\bm{x}}_i+{\bm{\eta}}_i),
\end{equation} 
\end{linenomath*}
where $\mathcal{L}_p$ refers to the objective function in (\ref{PAIB2}). The obtained perturbation then will be clipped {by CLIP function} so that $\|\Beta_i\|_{\infty}<\epsilon$.

In practice, all unlearnable perturbations ${\bm{\eta}}_i$ will be transferred to the authorized user via a secure channel. To mitigate the transmission overhead, we adopt the class-wise perturbation by taking the expectation of those ${\bm{\eta}}_i$ of $\tilde{\bm{x}}_i$'s from the same class. Thus, we can obtain $C$ perturbations as follows: 
\begin{equation} \label{classwise}
{\bm{\eta}}_c = \frac{1}{|\tilde{\mathcal{D}}_{c}|}{\sum \limits_{\tilde{\bm{x}}_i \in \tilde{\mathcal{D}}_{c}} {\bm{\eta}}_i},
\end{equation} 
where $\tilde{\mathcal{D}}_{c}$ contains all watermarked images of the $c$-th class. 

The generation of unlearnable perturbation using our proposed method is summarized in Algorithm \ref{algorithm1}. We train the victim model $f$ over the protected dataset by minimizing the cross-entropy loss within $T$ epochs (lines 2-5). Upon having a temporary victim model, we then generate unlearnable perturbations for each watermarked image (lines 6-10). These perturbations will be summarized according to labels of watermarked images, resulting in class-wise unlearnable perturbations $\Beta_c$ ($c=1,...,C$) (lines 11-13). The above procedures will be repeated until the classification accuracy $AC$ of the victim model achieves a stop criterion $e$. 

{
\begin{algorithm}
    \caption{The Proposed Unlearnable Examples} 
    \label{algorithm1} 
    \textbf{Input}: watermarked images $\tilde{\mathcal{D}}$, stop criterion $e$, train step $T$, learning rates $l_\theta,l_\eta$\\
    \textbf{Output}: class-wise unlearnable perturbation $\Beta_c$ $(c=1,...,C)$\\
    \textbf{Initialize}: $\Beta_c = \bm{0} (c=1,...,C)$
    \begin{algorithmic}[1]
    \WHILE{$AC < e$} 
        \STATE \#\#\#\#\#\#\# training the victim model $f$ \#\#\#\#\#\#\#
        \FOR{$t=1,...,T$}
        \STATE update the parameters of the model $f$ over the dataset $\tilde{\mathcal{D}}$ with perturbation $\Beta$.
        \ENDFOR
        \STATE \#\#\# generating the class-wise perturbation \#\#\#
        \FOR{$\tilde{\bm{x}}_i \in \tilde{\mathcal{D}}$}
        \STATE calculate $\Beta_i$ according to equation (\ref{JEPG_I}).
        \STATE $\Beta_i = \text{CLIP}(\Beta_i,-\epsilon,\epsilon)$. 
        \ENDFOR
        \FOR{$c=1,...,C$}
        \STATE calculate $\Beta_c$ according to equation (\ref{classwise}).
        \ENDFOR
        \STATE calculate the classification accuracy $AC$ over $\tilde{\mathcal{D}}$ with $\Beta$.
        \ENDWHILE 
    \end{algorithmic}
\end{algorithm}
}

{Our core motivation posits that MI-minimization perturbations fundamentally disrupt the model's semantic information extraction from images. To validate this, we quantitatively measure the mutual information $I(\hat{X};\hat{Y})$ between poisoned inputs $\hat{X}$ and model outputs $\hat{Y}$ across diverse architectures.
The perturbations were generated through mutual information minimization on the ImageNet-100 dataset using ResNet18 as the victim model. Using the mutual information of normal models on clean data as the baseline, MI decreased by 97\% for ResNet18, 82\% for VGG19, 89\% for MobileNet, 83\% for GoogleNet, and 97\% for Dense121, respectively.
All models exhibit drastic MI reduction on poisoned data compared with clean data. Consequently, despite perturbations being optimized only for ResNet18, MI values remain uniformly low across all architectures.
}

\subsection{Dual Watermark Extraction\label{dualwatermarkextraction}}
Due to the interference of the unlearnable perturbation $\bm{\eta}$, we use a dual decoder $\bm{D}_d$, the same network structure as the watermark decoder $\bm{D}$, to extract the same watermarks from the watermarked images with the unlearnable perturbation. The watermark extracted by $\bm{D}_d$ naturally is measured by 
\begin{linenomath*}
\begin{equation}
\begin{split}
\mathcal{L}_2&=\text{BCE}(\bm{m},\bm{m}_d'), ~\bm{m}_d'=\bm{D}_d(J(\tilde{\bm{x}}_i+\bm{\eta}_i);\bm{\theta}_{\bm{D}_d}),
\end{split}
\end{equation}
\end{linenomath*}
where $\bm{m}_d'$ is the watermark extracted by the dual decoder $\bm{D}_d$, $\bm{\eta}_i$ is the unlearnable perturbation, and $\bm{\theta}_{\bm{D}_d}$ are parameters of the dual decoder.
Now the complete watermark loss function consists of $\mathcal{L}_1$ and $\mathcal{L}_2$, and we designed and compared different optimization strategies in Section \ref{AblationStudy}.

\subsection{{Reversibility of Unlearnable Example}}

{
Perturbation plays a crucial role in protecting semantic information. After removing the perturbation, the image only contains watermark information, and the model's performance trained on the watermark image is normal.
Dataset learnability is reversible for authorized users, which is achieved through class-wise perturbation design. For each class $c\in\{1,...,C\}$, a universal perturbation $\bm{\eta}_c$ is generated and applied uniformly to all watermarked images $\tilde{\bm{x}}_i$ belonging to class $c$. Formally, the protected image is constructed as $\hat{\bm{x}}_i=\tilde{\bm{x}}_i+\bm{\eta}_c$, where $c=y_i$. The class-wise perturbations $\{\bm{\eta}_c\}^C_{c=1}$ are transmitted to authorized users via a secure channel. Authorized users possessing both the perturbations and the corresponding class labels $y_i$ can recover the original watermarked images by removing the perturbation: $\tilde{\bm{x}}_i=\hat{\bm{x}}_i-\bm{\eta}_c$.
}

\begin{table*}\caption{The performance of proposed copyright protection mechanism. The values in the method column's bracket are the intensity of unlearnable perturbations. The results are formatted as unlearnable accuracy (recovery accuracy) in the semantic information protection part and BER(PSNR) in the task of ownership verification. The notation ``-- --" means this measure does not apply to this method.}
\vspace{-0.5em}
\resizebox{\textwidth}{!}{%
\begin{tabular}{l|c|cccccc|cc}
\toprule[1pt]
\hline
\rule{0pt}{8pt}\multirow{2}{*}{Dataset} &\multirow{2}{*}{Method} & \multicolumn{6}{c|}{Semantic Information Protection (\%)}    & \multicolumn{2}{c}{Ownership Verification}                                \\ \cline{3-10}
\rule{0pt}{8pt}& & ResNet18       & VGG19          & MobileNetV2    & GoogLeNet      & Dense121  & Avg. & BER w/ & BER w/o      \\ \hline
\rule{0pt}{8pt}\multirow{6}{*}{ImageNet-100} & Original  & 71.39          & 65.09          & 76.80          & 77.13          & 73.50   & 72.78    & N/A      & N/A   \\
& GEAA\cite{zhao2022guided}         & 40.68(67.42)   & 41.88(63.30)   & 40.32(62.58)   & 47.76(71.64)   & 41.12(68.38) & 42.35(66.66)  & -- -- (9.23)           & 26.46\%(34.77)               \\
& Adv-watermark\cite{jia2020adv} & 59.94(-- --) & 46.88(-- --) & 60.94(-- --) & 61.84(-- --) & 61.30(-- --) & 58.18(-- --)            & -- -- (28.44)          & -- -- (-- --)         \\
& Ours(4)                               & {8.96(69.72)}    & {2.06(65.69)}    & {6.67(76.32)}    & {8.50(76.41)}    & {7.46(73.04)}    & {6.76(72.23)}    & \textbf{0.16\%}(41.25)             & \textbf{0.22\%}(48.36)\\
& Ours(8)                 & 5.46(69.98)    & 2.64(66.44)    & 1.28(75.38)    & 3.00(75.44)    & 6.46(72.36)   &3.77(71.92) & 0.27\%(35.54)             & 0.26\%(48.35)              \\ 
& Ours(16)                & 2.03(69.42)    &\textbf{ 1.30}(65.92)    & \textbf{1.16}(73.12)    & \textbf{1.52}(75.66)    & \textbf{1.98}(72.60)  &\textbf{1.60}(71.34) & 0.32\%(30.74)             & 0.35\%(48.21)                  \\ \hline \hline
\rule{0pt}{8pt}\multirow{2}{*}{{Pets}} & {Original} & {76.27} & {50.37
} & {70.79
} & {81.54
} & {75.73
} & {70.94
} & {N/A} & {N/A}\\
& {Ours(8)} & {2.70(73.36)} & {2.70(47.48)} & {2.70(70.67)} & {2.70(78.30)} & {2.70(73.02)} & {2.70(68.56)} & {0.25\%(37.07)} & {0.08\%(48.80)}\\ \hline \hline
\rule{0pt}{8pt}\multirow{3}{*}{CIFAR10} & Original  & 94.49 & 90.23 & 84.86 & 87.42 & 94.46 & 90.29 & N/A  & N/A  \\
& GEAA\cite{zhao2022guided}& 46.39(81.63)    & 45.18(77.71) & 41.98(64.98)       & 44.66(54.55)     & 49.54(72.84)  &   45.55(70.34)   & -- -- (16.21)    & 28.86\%(36.28)  \\

& Ours(4) & \textbf{10.50}(94.43)    & \textbf{11.68}(90.10) & \textbf{11.50}(84.26)       & \textbf{14.01}(86.96)     & \textbf{11.26}(94.05) & \textbf{11.79}(89.96)  & \textbf{2.03\%}(34.69) & \textbf{1.29\%}(38.27)\\ \hline
\bottomrule[1pt]
\end{tabular}}
\label{CopyrightProtection}
\vspace{-1.0em}
\end{table*}
\section{Experiments\label{Experiments}}
Our proposed method is a copyright protection mechanism that could effectively prevent the image dataset from unauthorized training (i.e., semantic information protection) and simultaneously avoid images with or without perturbation leakage by authorized users (i.e., ownership verification). To the best of our knowledge, our work is the first one that considers such a comprehensive copyright protection scenario. We select two recent studies closest to our scenario to assess the effectiveness of our proposed methodology: GEAA \cite{zhao2022guided} and Adv-watermark \cite{jia2020adv}. The GEAA considers recovered data leakage caused by authorized users and Adv-watermark adopts the visible watermarks which also function as an unlearnable perturbation. 
{The evaluation of our proposed method for tackling the above two threats is carried out on the ImageNet-100 dataset, covering the first 100 categories of ImageNet \cite{5206848}, CIFAR10 \cite{8612873}, and Oxford-IIIT Pets \cite{parkhi12a}.}

Ours' protected data generation includes three parts in order: 1) training the watermarking network including an encoder and a decoder, generating high-quality watermarked images; 2) unlearnable perturbation generation; 3) training the dual decoder to extract the watermark from the images combining the watermarked images and perturbations. 
The watermark performance and ablation study are detailed in Section \ref{OwnershipVerification} and \ref{AblationStudy}, and the unlearnable perturbation performance and robustness analysis are shown in Section \ref{SemanticInformationProtection}, \ref{PerformanceoftheProposedPoisoningAttack} and \ref{RobustnessAnalysis}.

\subsection{Semantic Information Protection\label{SemanticInformationProtection}}
For the task of semantic information protection, we train the different networks on the protected dataset and test on the clean test dataset. As shown in Table \ref{CopyrightProtection}, we can see that the accuracy of five considered models (the 6-th row of columns 3-8) is degraded closely aligning with the worst accuracy of a 100-class classification task, the $1\%$. Moreover, the values in parentheses are the accuracy of models trained over the images after removing the unlearnable perturbations. They are quite close to the ideal accuracy of models trained over clean images. In comparison, compared to the original accuracy, GEAA merely exhibited a modest reduction to $42.35\%$ on average and a $6\%$ decrease during authorized training (the 4-th row of columns 3-8). Adv-watermark also performs unsatisfactorily by merely degrading the original accuracy to $58.18\%$ (the 5-th row of columns 3-8). Since the watermarks (or say the unlearnable perturbations) generated by Adv-watermark are irreversible, the authorized training results are not presented. Note that our class-wise unlearnable perturbation is exactly reversible and the slight difference in accuracy between the original model and the authorized training model is caused by the embedded watermarks. 

Besides, it is important to mention that all five models are trained using the same unlearnable dataset obtained from ResNet18 and their accuracy is consistently degraded. This reflects the powerful generalizability of our proposed method, ensuring that we can successfully protect the semantic information of images in real scenarios where the models of authorized users are generally unknown. 

The 7-8 rows in Table \ref{CopyrightProtection} show the performance of the semantic information protection using unlearnable perturbation with $L_\infty$ magnitude 8 pixels and 16 pixels, respectively. We can derive the same conclusion as the case of 4 pixels $L_\infty$ magnitude, which reflects that our method does not rely on specific choices of perturbation magnitudes. In a nutshell, the results on ImageNet-100 support the effectiveness of our method of semantic information protection. 
{Besides, we evaluate the proposed method on Pets, the 9-10 rows of Table \ref{CopyrightProtection}, and five models trained on protected Pets data all achieve near-random accuracy 2.7\%. For authorized users, we recover 97\% of the original accuracy (68.56\% vs. 70.94\%). A similar conclusion can be derived on the CIFAR10.}

In the comparison methods, GEAA \cite{zhao2022guided} is only tested on the CIFAR10 in the original paper, and we use the open-source code setting the same watermark capacity as ours to test on the ImageNet-100. 
{Images were resized to $224\times224$. We trained GEAA for 15 phases, with each phase containing 30 epochs (batch size=8). At each phase start, we reset the learning rate to 0.0001 and decayed it by 0.5 after 10 epochs (lower boundary constraint=10).  
For Adv-watermark \cite{jia2020adv}, we attacked ResNet-101 using the Berkeley University badge (scale=1/4) with crossover probability=0.9, iterations=3, and population size=50.
we attack the 130000 images and run this experiment in 10 days with CPU I9-13900K + GPU 4090 configuration.} Because the watermark picture is too large, it cannot be implemented in CIFAR10. More details about training are discussed in Section \ref{PerformanceoftheProposedPoisoningAttack}
.

\subsection{Ownership Verification\label{OwnershipVerification}}
Now we evaluate the effectiveness of our method on the task of ownership verification. We use the Bit Error Rate (BER) metric to assess watermark extraction efficiency. Our work focuses on two types of ownership verification, i.e., the watermark extraction with unlearnable perturbations (denoted by BER w/) and that without perturbations (denoted by BER w/o). Also, we evaluated the quality of the protected images by using the Peak Signal to Noise Ratio (PSNR). It is noteworthy that the watermark is tested with JPEG compression with a widely used quality factor of 75 to simulate the real application scenario. It is worth noting that we do not assume the protector knows whether watermarked images have unlearnable perturbations. If either extractor can extract the information correctly, we achieve copyright verification. 

Table \ref{CopyrightProtection} records the performance of ownership verification of images in ImageNet-100, where a random 64-bit message is embedded into each image with a resolution of $256\times256$, respectively. As can be seen from the last two columns, our proposed method attains a BER of 0.16\% and 0.22\% (the 6-th row of the last two columns) with and without unlearnable perturbation, respectively, while maintaining high image quality. In comparison, GEAA performs unsatisfactorily with BER w/o of 26.46\% (the 4-th row of the last column). Notably, the GEAA framework sacrifices visual quality to protect the semantic information of the dataset. Adv-watermark, being a visible watermark attack method, is evaluated solely based on PSNR. 
{With the same setting, we test the watermark robustness on Pets. With perturbations present, BER is 0.25\%. Without perturbations, the BER improves to 0.08\%.}
The effectiveness of watermark extraction is also corroborated by results on CIFAR10 (bottom part of the table), where the 32-bit message was embedded in each $32\times32$ size image, respectively. 

\begin{figure}
  \centering
  \includegraphics[width=0.4\textwidth]{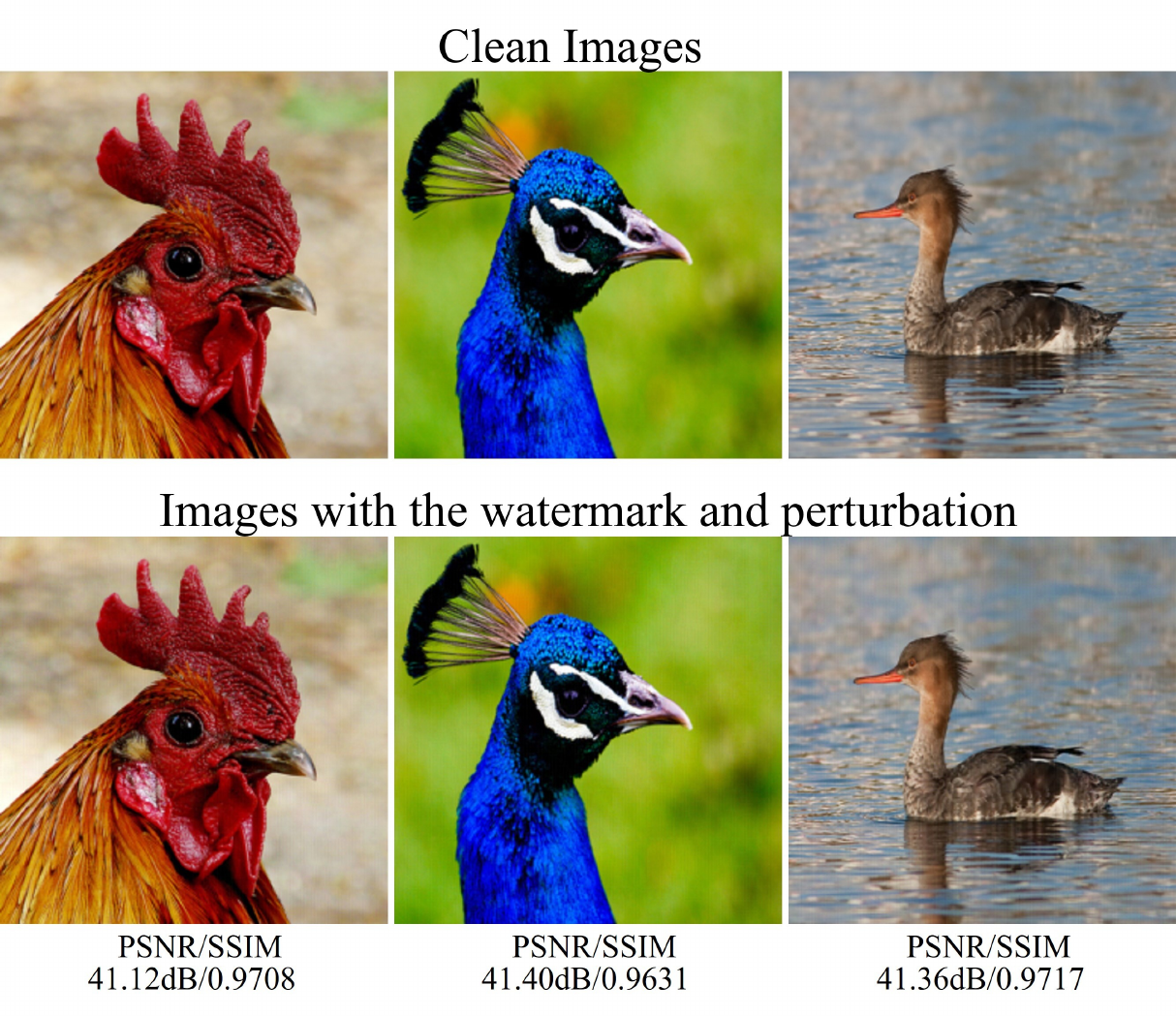}
  \vspace{-0.5em}
  \caption{Clean images (Top) VS. Images with the watermark and perturbation (Bottom).}
  \label{visual_protect_author_image}
\vspace{-0.5em}
\end{figure}

From the last two columns, we can observe that the magnitudes of unlearnable perturbations do not impact the performance of watermark extraction, which provides us with a flexible selection of perturbation magnitudes. Some visualization results are presented in Fig. \ref{visual_protect_author_image}. It is evident that the integrated watermark and perturbation have a minimal impact on the visual quality of images, which highlights the practicality of our method.
There are more visualization results in Fig. \ref{fig7}, where the images from left to right include clean images, images with perturbation generated by Adv-watermark, and images with and without perturbation by GEAA and Ours. There are PSNR and SSIM values below each image.

\subsection{Real-World Validation}
{To test the practicality in real-world scenarios, we conducted experiments on commercial API platforms for deep learning model training. In this setting, the model architectures and training strategies are unknown, simulating a realistic environment where unauthorized users might attempt to exploit the protected dataset. 
Following established methodology \cite{zhang2023unlearnable}, we test robustness on Baidu EasyDL, a widely used commercial deep learning platform. Using "fastest training" mode configurations, we evaluate three protected datasets: ImageNet10, CIFAR10, and Pets, and results are shown in Table \ref{table-commercialAPI}. Our method degrades accuracy to 13.80-18.14\%, significantly below the unprotected performance.} 

\begin{table}
\caption{{The test accuracy (\%) of models trained by EasyDL platforms. The training configuration on the platform was set to “fastest training”.}}
\vspace{-1em}
\centering
\resizebox{0.3\textwidth}{!}{
\begin{tabular}{cccc}
\toprule[1pt]
\hline
\rule{0pt}{8pt}{Data} & CIFAR10 & ImageNet10 & Pets \\ \hline  
\rule{0pt}{8pt}Clean & 92.20 & 77.00 & 67.03 \\ 
Emax\cite{fowl2021adversarial} & 18.88 & 20.60 & 20.58 \\ 
Emin\cite{huang2021unlearnable} & 19.47 & 23.80 & 37.24 \\ 
Ours & \textbf{16.33} & \textbf{13.80} & \textbf{18.14} \\ \hline
\bottomrule[1pt]
\end{tabular}}
\label{table-commercialAPI}
\vspace{-1em}
\end{table}

\subsection{Ablation Study}\label{AblationStudy}
\begin{table}
\caption{Watermark performance.}\label{table:watermark}
\vspace{-1em}
\centering
\resizebox{0.4\textwidth}{!}{
\begin{tabular}{lc|cccccc}
\toprule[1pt]
\hline
\rule{0pt}{8pt}Method &$\epsilon$  & BER w/ & BER w/o & PSNR & SSIM \\ \hline
\rule{0pt}{8pt}\textbf{single} & 8 & 0.98\% & 0.50\% & \textbf{48.99} & \textbf{0.9950}\\
\textbf{dual} & 8 & \textbf{0.38\%} & \textbf{0.19\%} & 45.82 & 0.9897 \\
\textbf{$\text{dual}_f$} & 8 & 0.51\% & 0.33\% & 48.21 & 0.9938 \\ \hline
\rule{0pt}{8pt}\textbf{single}  & 16 & 2.90\% & 0.64\% & \textbf{48.99} & \textbf{0.9950} \\
\textbf{dual} & 16  & 1.33\% & \textbf{0.34\%} & 45.82 & 0.9897 \\
\textbf{$\text{dual}_f$} & 16 & \textbf{0.81\%} & \textbf{0.34\%} & 48.21 & 0.9938 \\ \hline
\bottomrule[1pt]
\end{tabular}}
\vspace{-1.0em}
\end{table}

To empirically demonstrate the disruptive impact of embedding unlearnable perturbations on the watermark extraction process, we consider three methods of perturbation and watermark embedding and test on the validation set of ImageNet-100 containing 5000 images. The first one, denoted as \textbf{single}, is a straightforward combination of embedding the watermark followed by the unlearnable perturbation. The watermark extraction process before and after removing the perturbation uses the same extractor. Take the 2-nd row of Table \ref{table:watermark} as an example, we can see that extraction performances are highly impacted by the perturbation by increasing the BER from 0.50\% to 0.98\%. The second considered method is our proposed dual extractor, referred to as \textbf{dual}, which involves two independent extractors. Compared with the \textbf{single} design, our proposed method significantly eliminates the impact of unlearnable perturbations, where the perturbations merely increase the BER from 0.19\% to 0.38\% (the 3-rd row). Even when the perturbation magnitudes intensify to 16, the impact remains relatively subtle, degrading from 0.34\% to 1.33\% (the 6-th row). We observe that the dual extractor design moderately deteriorates the image qualities. This is because the encoder of watermarks has to embed more information into the images so that another extractor can successfully retrieve watermarks undergoing the impact of perturbations. To prioritize high image quality, we can freeze the pre-trained watermark encoder module by merely training another extractor to retrieve watermarks from watermarked images with perturbations. We denote this strategy as \textbf{$\text{dual}_f$}. Compared with the \textbf{dual} design, the \textbf{$\text{dual}_f$} strategy has an improvement in the image quality from PSNR 45.82 to 48.21 (the 5-th column of rows 6-7) with a slight cost of BER degradation from 0.81\% to 1.33\% (the 3-rd column of rows 6-7).

\begin{figure}
\centering
\includegraphics[width=0.4\textwidth]{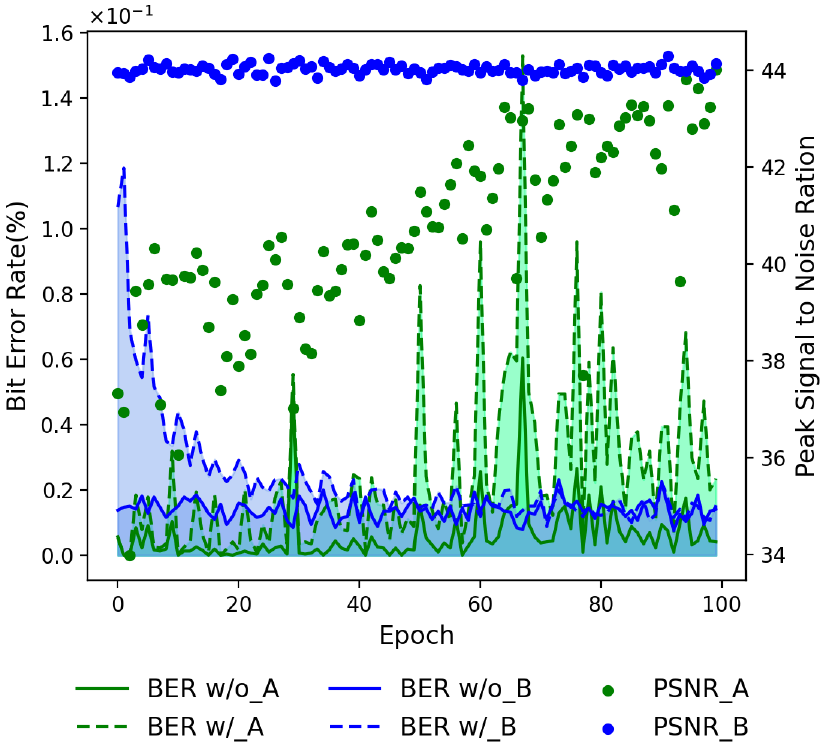}
\vspace{-0.5em}
\caption{{Performance comparison (BER$\downarrow$ and PSNR$\uparrow$) of two dual watermark extraction methods.} \textcolor[HTML]{008000}{A} and \textcolor{blue}{B} mean \textbf{dual} and $\textbf{dual}_f$ method respectively. BER w/o and BER w/ are the watermark robustness of images without and with unlearnable perturbation. Line-style stands for BER value referring to the left y-axis and point-style stands for PSNR referring to the right y-axis.}
\label{fig12}
\vspace{-0.5em}
\end{figure}

In our method, the embedding message is randomly generated for each batch in the training process, but the comparison method GEAA only embeds the pre-defined 10 kinds of messages. The training epoch is 100 and the batch size is 32, the best network parameters are good image quality and strong watermark robustness on the validation set. The training process is shown in Fig. \ref{fig12}. The training process of \textbf{dual} improves image quality but degrades robustness. In contrast, $\textbf{dual}_f$ trains the decoder in the case of freezing the encoder parameters, ensuring that the image quality is high when the watermark is strong robustness. The $\textbf{dual}_f$ method successfully extracts dual watermark information without compromising image quality.

To further prove the necessity and effectiveness of the proposed method, we consider a straightforward alternative by placing the watermark and perturbation in non-intersecting locations within the image. We split the images of CIFAR10 into two equal patches, and then embed watermarks and perturbations into the two patches, respectively. The average accuracy of the five models considered in our method increases from 11.45 to 43.73. The watermark capacity and robustness are also reduced. Those results support the effectiveness and the necessity of our proposed method.

\subsection{Performance of the Proposed Unlearnable Examples\label{PerformanceoftheProposedPoisoningAttack}}
Considering that our semantic information protection task requires a favorable availability attack with powerful generalizability, we propose a novel method to generate unlearnable perturbations by minimizing MI between the input image and the output of the victim model. In this section, let us discuss the performance of our proposed method alone.  
\begin{figure}
\centering
  \includegraphics[width=0.4\textwidth]{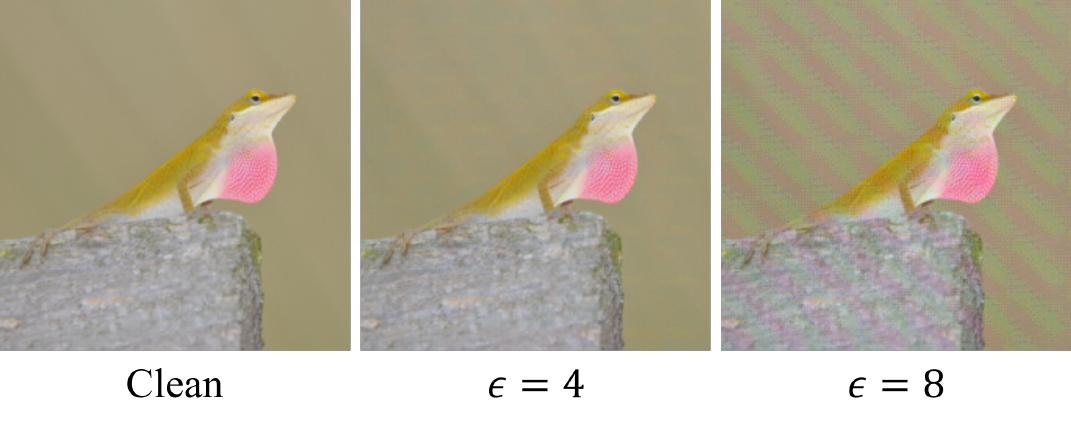}
 \vspace{-1em}
  \caption{Visualization of unlearnable images with different $\epsilon$}
  \label{fig5}
\vspace{-1em}
\end{figure}

\begin{figure*}
  \centering
  \includegraphics[width=0.8\textwidth]{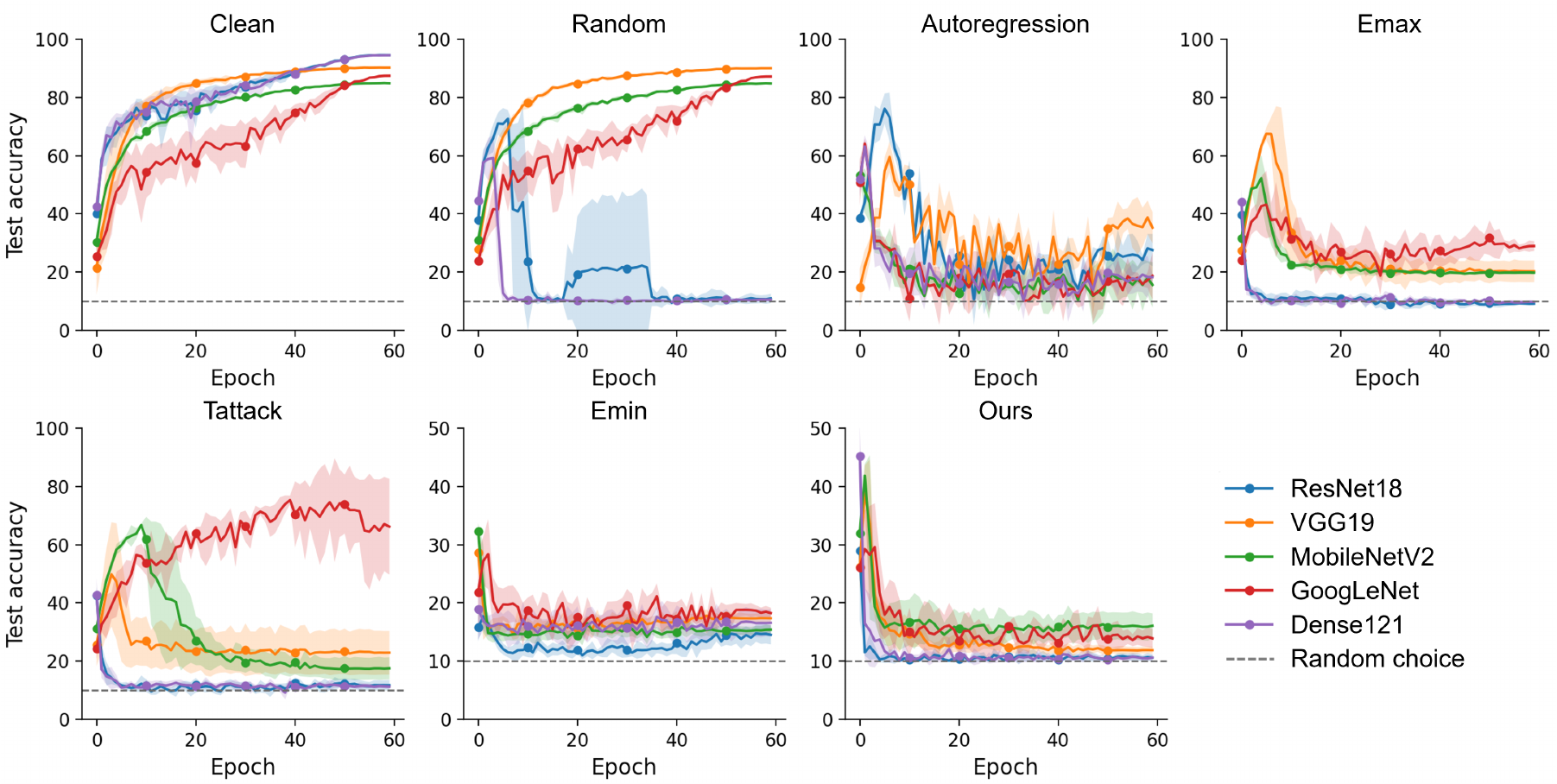}
  \caption{Training accuracy ($\downarrow$) of networks on protected CIFAR10 dataset using various availability attack methods. {Curves show mean accuracy over 6 experimental repetitions, with shading indicating standard deviation.}}
  \label{fig6}
\vspace{-0.5em}
\end{figure*}

In previous adversarial research, the commonly used noise constraint $\Vert\bm\eta_i\Vert_{\infty}<\epsilon=8$ does not guarantee imperceptibility to humans, as evident in our visualization results in Fig. \ref{fig5}. So we set $\epsilon=4$ in our experiments to provide high image quality. 
To test the transferability of noise, we choose five distinct network architectures: ResNet18 \cite{he2016deep}, VGG19 \cite{simonyan2014very}, MobileNetV2 \cite{sandler2018mobilenetv2}, GoogLeNet \cite{szegedy2015going}, and Dense121 \cite{huang2017densely}.
We utilize cross-entropy loss as the criterion, employing SGD with a momentum of 0.9 and a weight decay of 0.0005 as the optimizer. Combined with the CosineAnnealingLR decay scheduler, the initial learning rate is 0.01 and gradually declines to 0. The batch size is 128 for CIFAR10 and 32 for ImageNet. Data augmentation methods include horizontal flipping and random cropping. The training epochs are set to 60 for CIFAR10 and 100 for ImageNet-100.

To assess the effectiveness of our method, we compare it with multiple state-of-the-art availability attacks, including Emax \cite{fowl2021adversarial}, Emin \cite{huang2021unlearnable}, Tattack \cite{fowl2021adversarial}, Linear \cite{peng2022learnability}, Conv \cite{peng2022learnability}, Autoregressive \cite{sandoval2022autoregressive}, and Gradient alignment \cite{fowl2021preventing}. As presented in Tables \ref{table1}, the experimental results include the mean and standard deviation of the test accuracy over 6 times repeats. The best scores are highlighted in bold. The results reveal that our method attains the best average score across five models, surpassing the second-best by 6 points on CIFAR10 (the last column). Especially, the accuracy of five different models is quite consistent, which reveals the favorable generalizability of our proposed attack method. Similar trends are mirrored in the results for ImageNet-10 and ImageNet-100. Collectively, these findings underscore the efficacy of our method in safeguarding data copyright.

In Fig. \ref{fig6}, the training process of different methods on CIFAR10 with $\epsilon=4$ perturbations are shown, with the shadow indicating the standard deviation and signifying training stability. Notably, weak models with lower accuracy in original data training, such as MobileNetV2, and GoogLeNet, exhibited robust to random noise. Perturbations generated by Tattack \cite{fowl2021adversarial}, Emax \cite{fowl2021adversarial}, and Autoregression \cite{sandoval2022autoregressive} demonstrated poor transferability on weak models. Compared to Emin \cite{huang2021unlearnable}, our approach yields lower accuracy and showcases greater performance stability.

\begin{table}
\vspace{-1.0em}
\centering
\caption{Data augmentation on the CIFAR10. SIP stands for semantic information protection, and OV for ownership verification. Results marked with * refer to the original paper.}
\vspace{-0.5em}
\label{table5}\resizebox{0.45\textwidth}{!}{%
\begin{tabular}{ll|cccc}
\toprule[1pt]
\hline
\rule{0pt}{8pt}Task & Method & Gaussian noise & Cutmix & Cutout &Mixup\\ \hline
\rule{0pt}{8pt}\multirow{6}{*}{SIP} &Emax\cite{fowl2021adversarial} & $\text{10.88}_{\pm \text{0.35}}$ & $\text{12.57}_{\pm \text{0.58}}$  & $\text{11.90}_{\pm \text{0.33}}$ & $\text{17.26}_{\pm \text{0.71}}$\\
&Emin\cite{huang2021unlearnable} & $\text{10.62}_{\pm \text{0.17}}$ & $\textbf{10.54}_{\pm \text{0.04}}$  & $\text{10.91}_{\pm \text{0.02}}$ & $\text{18.14}_{\pm \text{0.92}}$ \\
&Tattack\cite{fowl2021adversarial} & $\text{14.78}_{\pm \text{1.04}}$ & $\text{12.18}_{\pm \text{0.63}}$  & $\text{12.06}_{\pm \text{2.28}}$ & $\text{13.42}_{\pm \text{1.69}}$ \\
&Linear*\cite{peng2022learnability} & 19.83 & 20.72  & 26.28 & 43.88 \\ 
&Conv*\cite{peng2022learnability} & 21.28 & 18.60  & 23.04 & 36.97 \\ 
&Ours & $\textbf{10.37}_{\pm \text{0.27}}$ & $\text{11.04}_{\pm \text{1.64}}$  & $\textbf{10.52}_{\pm \text{0.19}}$ & $\textbf{12.88}_{\pm \text{0.55}}$ \\ \hline
\rule{0pt}{8pt}OV &Ours &0.06\% &0.05\% &0.03\% &0.04\% \\

\hline
\bottomrule[1pt]
\end{tabular}}
\vspace{-1.0em}
\end{table}

\begin{table*}
\caption{Generalizability comparison of unlearnable perturbation. Results marked with * refer to the original paper.}
\centering
\resizebox{0.8\textwidth}{!}
{%
\begin{tabular}{c|l|cccccc}
\toprule[1pt]
\hline
\rule{0pt}{8pt}Dataset & Method  & ResNet18 & VGG19 & MobilenetV2 & GoogLeNet & Dense121 & Avg.\\ 
\hline
\rule{0pt}{8pt}\multirow{9}{*}{CIFAR10} & Clean       & $\text{94.49}_{\pm \text{0.09}}$          & $\text{90.23}_{\pm \text{0.36}}$          & $\text{84.86}_{\pm \text{0.37}}$          & $\text{87.42}_{\pm \text{0.26}}$         & $\text{94.46}_{\pm \text{0.16}}$ & 90.29          \\
&Emax\cite{fowl2021adversarial}                    & $\textbf{9.20}_{\pm \text{0.63}}$                          & $\text{20.29}_{\pm \text{3.70}}$                     & $\text{19.80}_{\pm \text{0.65}}$                            & $\text{28.97}_{\pm \text{1.66}}$                         & $\textbf{9.82}_{\pm \text{0.65}}$                         & 17.62                    \\
&Emin\cite{huang2021unlearnable}    & $\text{16.51}_{\pm \text{2.39}}$                        & $\text{18.56}_{\pm \text{0.81}}$                     & $\text{16.94}_{\pm \text{1.16}}$                           & $\text{21.73}_{\pm \text{0.75}}$                         & $\text{15.37}_{\pm \text{1.72}}$                        & 17.82                    \\
&Tattack\cite{fowl2021adversarial}   & $\text{11.70}_{\pm \text{1.83}}$                         & $\text{22.91}_{\pm \text{7.52}}$                     & $\text{17.53}_{\pm \text{3.80}}$                           & $\text{66.20}_{\pm \text{16.46}}$                          & $\text{11.28}_{\pm \text{0.72}}$                        & 25.92                    \\
&Tattack(sample-wise)*\cite{fowl2021adversarial} & $\text{23.79}_{\pm \text{0.27}}$                         & $\text{28.49}_{\pm \text{0.65}}$                     & $\text{20.53}_{\pm \text{0.02}}$                           & $\text{23.05}_{\pm \text{0.16}}$                          & $\text{20.90}_{\pm \text{0.22}}$                        & 23.35                    \\
&Autoregressive\cite{sandoval2022autoregressive}  & $\text{25.98}_{\pm \text{1.20}}$        & $\text{37.52}_{\pm \text{0.93}}$                  & $\text{17.88}_{\pm \text{1.51}}$              & $\text{18.58}_{\pm \text{1.73}}$                    & $\text{18.12}_{\pm \text{1.34}}$                        & 23.56                    \\
&Gradient alignment*\cite{fowl2021preventing}               & $\text{68.15}_{\pm \text{0.55}}$                        & $\text{66.58}_{\pm \text{0.58}}$                      & $\text{66.71}_{\pm \text{0.60}}$                           & $\text{71.02}_{\pm \text{0.15}}$                         & $\text{70.12}_{\pm \text{0.12}}$                        & 68.52                    \\
&Ours          & $\text{10.26}_{\pm \text{0.15}}$                        & $\textbf{11.74}_{\pm \text{0.23}}$                      & $\textbf{11.27}_{\pm \text{0.45}}$                           & $\textbf{13.51}_{\pm \text{0.35}}$                         & $\text{10.45}_{\pm \text{0.40}}$                        & \textbf{11.45}                     \\ \hline
\rule{0pt}{8pt}\multirow{6}{*}{ImageNet-10} &Clean       & $\text{72.77}_{\pm \text{0.93}}$          & $\text{67.60}_{\pm \text{0.60}}$          & $\text{72.37}_{\pm \text{1.02}}$          & $\text{74.47}_{\pm \text{1.56}}$         & $\text{73.67}_{\pm \text{1.47}}$ & 72.18          \\
&Emax\cite{fowl2021adversarial} & $\text{26.37}_{\pm \text{0.53}}$  & $\text{20.93}_{\pm \text{0.36}}$   & $\text{24.20}_{\pm \text{0.82}}$  & $\text{27.30}_{\pm \text{0.82}}$  & $\text{35.90}_{\pm \text{1.72}}$  & 26.94                    \\
&Emin\cite{huang2021unlearnable} & $\text{17.23}_{\pm \text{0.37}}$  & $\textbf{10.73}_{\pm \text{0.51}}$  & $\textbf{14.77}_{\pm \text{0.99}}$  & $\text{22.33}_{\pm \text{1.91}}$ & $\text{42.93}_{\pm\text{2.93}}$  & 21.60                    \\
&Tattack\cite{fowl2021adversarial} & $\text{19.93}_{\pm \text{2.89}}$  & $\text{21.07}_{\pm \text{0.68}}$  & $\text{17.50}_{\pm \text{2.20}}$  & $\text{23.97}_{\pm \text{1.17}}$  & $\text{29.93}_{\pm \text{2.23}}$  & 22.48                    \\
&Ours & $\textbf{13.50}_{\pm \text{0.46}}$  & $\text{20.50}_{\pm \text{1.61}}$  & $\text{16.63}_{\pm \text{2.05}}$  & $\textbf{17.60}_{\pm \text{1.65}}$  & $\textbf{21.10}_{\pm \text{2.51}}$  & \textbf{17.87}                     \\ \hline
\rule{0pt}{8pt}\multirow{6}{*}{ImageNet-100}  &Clean & $\text{71.39}_{\pm \text{0.31}}$          & $\text{65.09}_{\pm \text{0.35}}$          & $\text{76.80}_{\pm \text{0.39}}$          & $\text{77.13}_{\pm \text{0.07}}$         & $\text{73.50}_{\pm \text{0.33}}$ & 72.78          \\
&Emax\cite{fowl2021adversarial} & $\text{11.37}_{\pm \text{0.62}}$  & $\text{4.19}_{\pm \text{2.26}}$   & $\text{8.65}_{\pm \text{0.13}}$  & $\text{12.15}_{\pm \text{0.23}}$  & $\text{12.30}_{\pm \text{0.60}}$  & 9.73                    \\
&Emin\cite{huang2021unlearnable} & $\text{9.09}_{\pm \text{0.35}}$  & $\text{2.55}_{\pm \text{2.17}}$  & $\textbf{2.65}_{\pm \text{1.37}}$  & $\text{1.52}_{\pm \text{0.17}}$ & $\text{24.98}_{\pm \text{1.48}}$  & 8.16                    \\
&Tattack\cite{fowl2021adversarial} & $\textbf{8.45}_{\pm \text{0.26}}$  & $\text{4.90}_{\pm \text{0.28}}$  & $\text{6.61}_{\pm \text{0.05}}$  & $\text{9.59}_{\pm \text{0.27}}$  & $\textbf{10.75}_{\pm \text{0.53}}$  & 8.06                    \\
&Ours & $\text{8.75}_{\pm \text{0.05}}$  & $\textbf{1.11}_{\pm \text{0.05}}$  & $\text{6.62}_{\pm\text{0.89}}$  & $\textbf{0.94}_{\pm\text{0.33}} $  &$\text{14.39}_{\pm\text{0.51}}$  & \textbf{6.36}                     \\ \hline
\bottomrule[1pt]
\end{tabular}}
\label{table1}
\vspace{-0.5em}
\end{table*}

\subsection{Robustness Analysis\label{RobustnessAnalysis}}
We address the issue of data compression in cloud-based data repositories and the susceptibility of protected data to attacks by data thieves. As a consequence, it becomes crucial to assess the robustness of protected data.

\subsubsection{JPEG Compression}
For watermark, we report the results of compression of watermarked images in Table \ref{CopyrightProtection}, and our BERs are all under 0.35\%. For perturbation,
our experiments unveiled the vulnerability of universal adversarial perturbations generated by state-of-the-art methods of availability attack when exposed to JPEG compression. Results are presented in Table \ref{table4}. Models trained on original images subjected to JPEG75 compression maintained stable accuracy levels (the 2-nd row). However, the accuracy of models trained on unlearnable examples produced by Emin, Emax, Tattack, and our method exhibited an average improvement of 38\% upon compression (3-6 rows of the last column). This counteracted the effect of original availability attacks.

In equation (\ref{PAIB2}), we referred to the open-source code from \cite{wu2022robust} for the implementation of both standard JPEG and differentiable JPEG algorithms. The quality factor (QF) for the differentiable JPEG in the perturbations optimization process is set to 75. We assessed their transferability using various QFs for the standard JPEG. 
The outcomes in Table \ref{table4} reveal a substantial enhancement in our method's robustness against JPEG compression when utilizing differentiable JPEG. The accuracy is reduced from 63\% to 39\% (the last column of 6-7 rows). Moreover, no significant variation is observed in the effect of different QFs or no compression (the 8-9 rows).

\begin{figure*}
  \centering
  \includegraphics[width=0.8\textwidth]{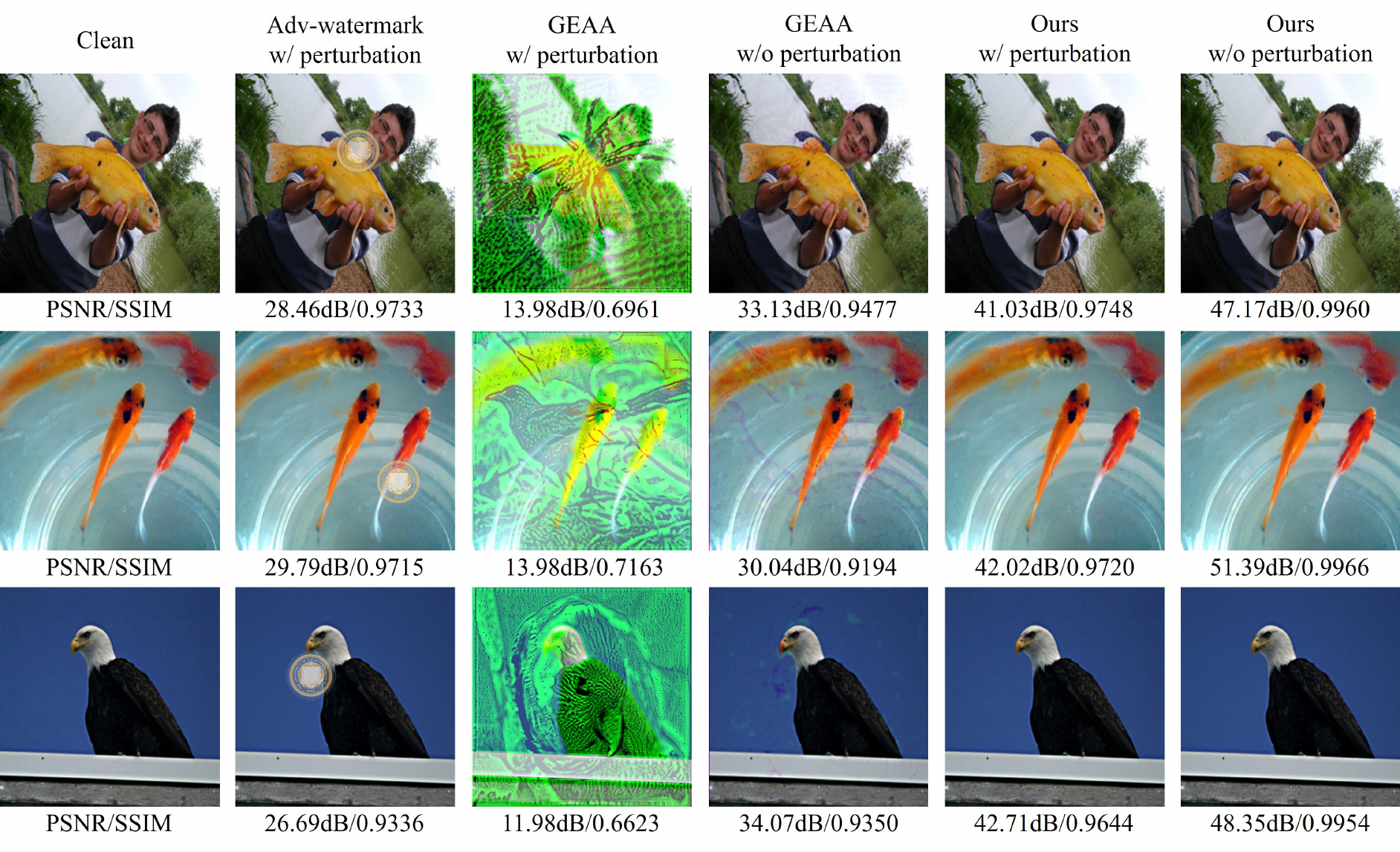}
  \vspace{-0.5em}
  \caption{Visualization of images with and without perturbation generated by different methods.}
  \label{fig7}
\vspace{-0.5em}
\end{figure*}

\begin{table*}
\caption{Effect of JPEG compression on CIFAR10 protection. QF is the JPEG quality factor, and distortion indicates the average change of image pixels.}
\vspace{-0.5em}
\centering
\label{table4}
\resizebox{0.75\textwidth}{!}{%
\begin{tabular}{l|cc|cccccc}
\toprule[1pt]
\hline
\rule{0pt}{8pt}Method & QF & distortion & ResNet18 & VGG19 & MobilenetV2 & GoogLeNet & Dense121 & Avg. \\ \hline
\rule{0pt}{8pt}Clean   &75 &5.97 & $\text{92.14}_{\pm\text{0.15}}$                        & $\text{88.61}_{\pm\text{0.08}}$                     & $\text{83.13}_{\pm\text{0.35}}$                           & $\text{85.52}_{\pm\text{0.16}}$                         & $\text{92.28}_{\pm\text{0.11}}$                        & 88.34 \\ \hline
\rule{0pt}{8pt}Emax\cite{fowl2021adversarial}    &75  &6.44  & $\text{46.35}_{\pm\text{ 0.25}}$ & $\text{53.07}_{\pm \text{1.72}}$ & $\text{46.82}_{\pm \text{1.38}}$ & $\text{75.07}_{\pm \text{8.43}}$ & $\text{43.26}_{\pm \text{0.64}}$  & 52.91 \\
Emin\cite{huang2021unlearnable}    &75     &6.51  & $\text{56.62}_{\pm \text{0.69}}$                        & $\text{45.12}_{\pm \text{1.00}}$                     & $\text{39.77}_{\pm \text{0.80}}$                           & $\text{51.77}_{\pm \text{0.93}}$                         & $\text{53.41}_{\pm \text{0.95}}$                        & 49.34                    \\
Tattack\cite{fowl2021adversarial}    &75  &6.46 & $\text{63.12}_{\pm \text{0.48}}$ & $\text{60.26}_{\pm \text{22.50}}$ & $\text{71.14}_{\pm \text{4.90}}$ & $\text{84.26}_{\pm\text{0.17}}$ & $\text{60.58}_{\pm \text{0.85}}$ & 67.87                    \\
Ours    &75    &6.62  & $\text{69.24}_{\pm \text{0.43}}$                        & $\text{62.60}_{\pm \text{0.86}}$                      & $\text{53.70}_{\pm \text{0.80}}$                           & $\text{64.87}_{\pm \text{0.54}}$                         & $\text{67.26}_{\pm \text{1.25}}$                        & 63.53                     \\ 
Ours+JPEG    &75   &\textbf{6.38}     & $\textbf{35.25}_{\pm \text{1.06}}$                        & $\textbf{38.80}_{\pm \text{13.08}}$                      & $\textbf{39.09}_{\pm \text{0.96}}$                           & $\textbf{50.57}_{\pm \text{1.13}}$                         & $\textbf{31.88}_{\pm \text{1.34}}$                        & \textbf{39.12}                     \\ \hline
\rule{0pt}{8pt}Ours+JPEG    &85   &5.62     & $\text{44.42}_{\pm \text{0.66}}$                        & $\text{39.76}_{\pm \text{1.17}}$                      & $\text{36.09}_{\pm \text{0.65}}$                           & $\text{48.74}_{\pm \text{0.92}}$                         & $\text{38.54}_{\pm \text{0.80}}$                        & 41.51                     \\
Ours+JPEG    &N/A   &0     & $\text{10.05}_{\pm \text{0.03}}$                        & $\text{12.24}_{\pm \text{0.77}}$                      & $\text{11.06}_{\pm \text{0.51}}$                           & $\text{15.60}_{\pm \text{1.54}}$                         & $\text{10.05}_{\pm \text{0.02}}$                        & 11.80                     \\\hline \bottomrule[1pt]
\end{tabular}}
\vspace{-0.5em}
\end{table*}

\subsubsection{Data Augmentation}
In real-world scenarios, data augmentation serves as a commonly employed technique to artificially expand data volumes. This approach proves beneficial for enhancing model performance by generating novel and distinct examples for training datasets. However, its utilization might counteract the impact of minor adversarial perturbations. To assess robustness, we examine various data augmentation methods for training a ResNet-18 model, encompassing Gaussian noise, Cutmix, Cutout, and Mixup. In the Gaussian noise attack, each image receives an additional noise layer with a random strength of $\epsilon=4$. All aforementioned experiments are conducted under standard augmentation settings, including horizontal flip and random crop. The results presented in Table \ref{table5} (the 2-7 rows) demonstrate the minimal influence of common data augmentation techniques on our method. We also pollute watermarks with these augmentation methods, and BERs are closed to 0 (the last row). 

\begin{table}
    \centering
    \caption{The robustness of adversarial training.}
    \vspace{-0.5em}
    \label{tab:adv_training}
    \resizebox{0.45\textwidth}{!}{%
    \begin{tabular}{lc|cccccc}
    \toprule[1pt]
    \hline
    \rule{0pt}{8pt}Method        &$\rho$       & ResNet18 & VGG19                        & MobileNetV2 & GoogLeNet & Dense121 & Avg    \\ \hline
\rule{0pt}{8pt}Emin\cite{huang2021unlearnable}         & 2       & 92.12    & \textbf{10.00}                        & 80.46       & 84.88     & 18.30    & 57.15  \\
Ours          & 2       & \textbf{16.21}    & \textbf{10.00}                        & \textbf{16.56}       & \textbf{14.50}     & \textbf{18.18}    & \textbf{15.09}  \\ \hline
\rule{0pt}{8pt}Emin\cite{huang2021unlearnable}         & 4       & 59.36    & \textbf{10.00}                        & 69.59       & 72.81     & 48.97    & 52.15  \\
Ours          & 4       & \textbf{39.79}    & \textbf{10.00}                        & \textbf{68.09}       & \textbf{69.22}     & \textbf{42.10}    & \textbf{45.84}  \\ \hline
\bottomrule[1pt]
\end{tabular}}
\vspace{-0.75em}
\end{table}

\subsubsection{The Robustness of Adversarial Perturbations}
At present, there are some defenses against availability attacks, including adversarial training\cite{fu2022robust}, adversarial augmentation\cite{qin2023learning}, and clean-unlearnable data mixture training.
We perform adversarial training on data protected by our method and the comparative methods Emin\cite{huang2021unlearnable}. Table \ref{tab:adv_training} illustrates that Emin is ineffective for adversarial training, particularly with adversarial perturbation $\rho=2$. In comparison, our proposed method can still effectively degrade the performance of models. As for the adversarial augmentation\cite{qin2023learning}, it has a limited impact on our method by simply improving the accuracy of ResNet18 from 10.26\% to 15.54\% on CIFAR10. For 10\% clean data combined with 90\% unlearnable data, Emin gets 77.18\% and ours is 66.84\%. It shows that our method still has an advantage.

\begin{table}
\centering
\caption{{Watermark Robustness Under Extended Distortions (BER$\downarrow$)}}
\vspace{-0.5em}
\label{table_WatermarkRobustness}
\resizebox{0.35\textwidth}{!}{%
\begin{tabular}{ccc}
\toprule[1pt]
\hline
\rule{0pt}{8pt}{Attack} & {BER (w/)} & {BER (w/o)} \\ \hline
\rule{0pt}{8pt}{No attack} & {0} & {0} \\
{Speckle (var=0.01)} & {0.03\%} & {0.08\%} \\
{Salt\&pepper (density=0.005)} & {0.49\%} & {0.33\%} \\
{Salt\&pepper (density=0.01)} & {1.50\%} & {1.06\%} \\
{Median filter {[}3, 3{]}} & {4.27\%} & {3.26\%} \\
{Gaussian ($\sigma$=0.001)} & {0.08\%} & {0.16\%} \\
{JPEG (QF=75)} & {0.28\%} & {0.34\%} \\
{Resize {[}256-128-256{]}} & {2.81\%} & {3.04\%} \\
{Resize {[}256-512-256{]}} & {0.01\%} & {0.01\%} \\ \hline
\bottomrule[1pt]
\end{tabular}}
\vspace{-0.5em}
\end{table}

{4) \textit{The Robustness of Watermark}: To comprehensively evaluate the robustness of the embedded watermark under diverse distortions, we conducted extensive experiments following \cite{10486948}. {Table \ref{table_WatermarkRobustness}} summarizes the BER values under extended distortions.
The watermark exhibits strong resilience to the speckle (var=0.01), salt\&pepper (density=0.005), Gaussian ($\sigma$=0.001), JPEG (QF=75), resize (256-512-256), with BER consistently below 0.35\% even with perturbations.}

\section{Conclusion\label{Conclusion}}
In this paper, we have proposed a data protection method that combines recoverable unlearnable examples and digital watermarking, which can realize the role of copyright verification and semantic information protection of datasets simultaneously. We propose a more effective method based on the theory of information entropy for availability attacks and verify it on multiple datasets. For digital watermarks, we propose a double watermark extraction method that can extract messages from two states of images. This method can solve the actual problems in data copyright protection, namely copyright verification against leaked data and authorization from protected datasets.
\section*{Acknowledgments}
This work was supported in part by the National Natural Science Foundation of China (Grant No. 62202009) and by the Macau Science and Technology Development Fund (Grant No. 0040/2023/ITP1, 0004/2023/RIB1).

\bibliographystyle{IEEEtran}
\bibliography{bibfile}
\vspace{-2.0em}
\begin{IEEEbiography}[{\includegraphics[width=1in,height=1.25in,clip,keepaspectratio]{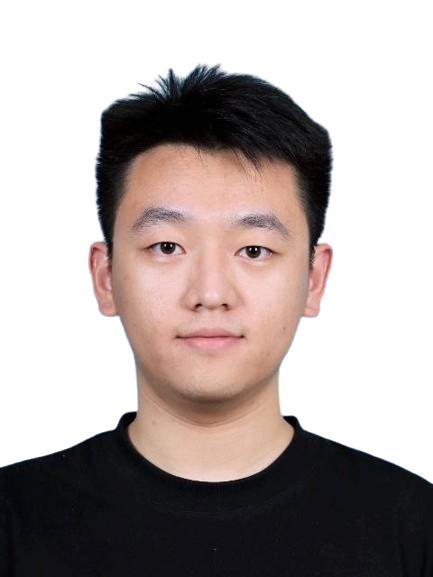}}]{Binze Wang} received the B.S. degree in the Chang'an University, Shaanxi, China, 2021 and M.S. from the Chinese Academy of Surveying and Mapping, Beijing, China, 2024. He is currently pursuing the Ph.D. degree in artificial intelligence at the Macau University of Science and Technology, Macau, China. His interests primarily lie in the security aspects of deep learning, especially adversarial attacks, and copyright protection.
\end{IEEEbiography}
\vspace{-2.0em}
\begin{IEEEbiography}[{\includegraphics[width=1in,height=1.25in,clip,keepaspectratio]{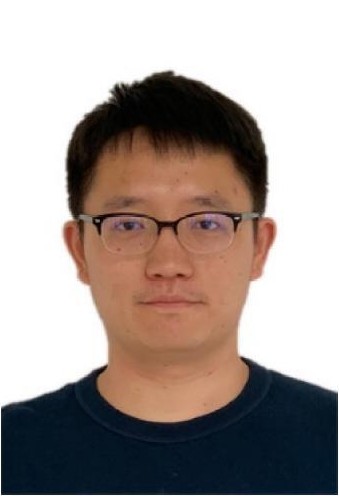}}]{Jinyu Tian} (Member, IEEE) received the B.S. and M.S. degrees in mathematics from Chongqing University, Chongqing, China, in 2014 and 2017, respectively. He got his Ph.D. degree in computer science from University of Macau, Macau, China, in 2022. He is currently an Assistant Professor at the School of Computer Science and Engineering, Macau University of Science and Technology. His current research interests include adversarial machine learning, security of deep learning, multimedia forensics, and subspace learning. He has published in several high-impact international journals and conferences, including IEEE TIP, TNNLS, TIFS, TSC, and AAAI, CVPR. He is also an active reviewer for numerous prestigious journals and conferences, such as IEEE TIP, TMM, ACM MM, CVPR, NeurIPS, ICML, etc.
\end{IEEEbiography}
\vspace{-2.0em}
\begin{IEEEbiography}[{\includegraphics[width=1in,height=1.25in,clip,keepaspectratio]{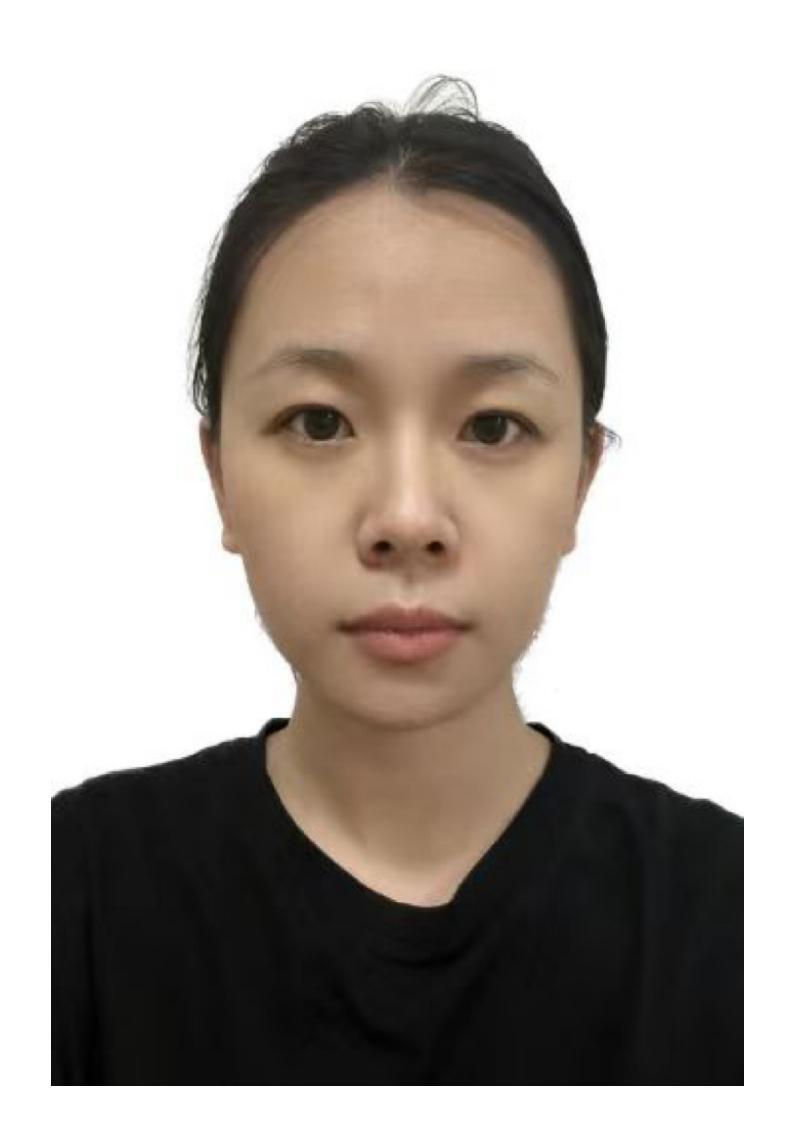}}]{Xingrun Wang}
received the M.S. degree from the School of Computer Science and Technology, Shandong University in 2019. She recieved the Ph.D. degree with the School of Computer Science and Engineering, Macau University of Science and Technology in 2024. Since 2025, she has been working at Foshan University. Her research interests include information security and protection, computer vision, and artificial intelligence. 
\end{IEEEbiography}
\begin{IEEEbiography}[{\includegraphics[width=1in,height=1.25in,clip,keepaspectratio]{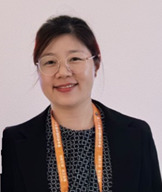}}]{Xiaochen Yuan}
(Senior Member, IEEE) received the Ph.D. degree in Software Engineering from the University of Macau in 2013. From 2014 to 2015, she was a postdoctoral fellow at the Department of Computer and Information Science of the University of Macau. From 2016 to 2021, she was an Assistant Professor and an Associate Professor at the Faculty of Information Technology of the Macau University of Science and Technology. Since 2021, she joined the Faculty of Applied Sciences of the Macao Polytechnic University, where she is currently an Associate Professor. Her research interests include Multimedia Forensics and Security, Digital Watermarking, AI Model Security, Quantum Watermarking, Remote Image Processing, and Deep Learning Techniques and Applications. She has published more than 70 SCIE-indexed scientific articles in refereed journals, such as IEEE TIFS, IEEE TII, IEEE TMI, IEEE TIM, IEEE TETC, IEEE JSTARS, etc. and she has been served as reviewers for top-tier journals in related areas, as well as the program committee member, session chair, and regional chair for ICSPS, ICSIP, ICSTE, etc. She is also a member of CCF, a member of CSIG, and a senior member of IEEE.
\end{IEEEbiography}
\vspace{-2.0em}
\begin{IEEEbiography}[{\includegraphics[width=1in,height=1.25in,clip,keepaspectratio]{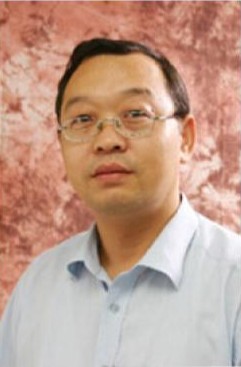}}]{Jianqing Li}
(Senior Member, IEEE)  received the Ph.D. degree in electronic engineering from the Beijing University of Posts and Telecommunications, Beijing, China, in 1999.
From 2000 to 2002, he was a Visiting Professor with Information and Communications University, Daejeon, South Korea. From 2002 to 2004, he was a Research Fellow with Nanyang Technological University, Singapore. He joined the Macau University of Science and Technology in August 2004. He is currently a Professor. His research interests include wireless networks, IoT, two-dimensional materials for photonics and fiber sensors.
\end{IEEEbiography}

\vfill
\end{document}